\documentclass[twocolumn,times]{aastex62}
\usepackage{graphicx}
\usepackage[flushleft]{threeparttable}
\usepackage{blindtext}
\usepackage{amsmath}
\usepackage{mathtools}
\usepackage{multirow}
\turnoffeditone

\begin{document}
\shortauthors{Jung et al.}
\def\nar{New Astron.}
\def\na{New Astron.}
\title{\Large \textbf{Texas Spectroscopic Search for Ly$\alpha$ Emission at the End of Reionization. III. the Ly$\alpha$ Equivalent-width Distribution and Ionized Structures at $z > 7$}}

\correspondingauthor{Intae Jung}
\email{intae.jung@nasa.gov}

\author[0000-0003-1187-4240]{Intae Jung}
\affil{Astrophysics Science Division, Goddard Space Flight Center, Greenbelt, MD 20771, USA}
\affil{Department of Physics, The Catholic University of America, Washington, DC 20064, USA }
\affil{Center for Research and Exploration in Space Science and Technology, NASA/GSFC, Greenbelt, MD 20771}
\affil{Department of Astronomy, The University of Texas at Austin, Austin, TX 78712, USA}

\author[0000-0001-8519-1130]{Steven L. Finkelstein}
\affil{Department of Astronomy, The University of Texas at Austin, Austin, TX 78712, USA}

\author[0000-0001-5414-5131]{Mark Dickinson}
\affil{NSF’s National Optical-Infrared Astronomy Research Laboratory, Tucson, AZ 85719, USA}

\author[0000-0001-6251-4988]{Taylor A. Hutchison}
\affil{Department of Physics and Astronomy, Texas A\&M University, College
Station, TX, 77843-4242 USA}
\affil{George P.\ and Cynthia Woods Mitchell Institute for
  Fundamental Physics and Astronomy, Texas A\&M University, College
  Station, TX, 77843-4242 USA}
  
\author[0000-0003-2366-8858]{Rebecca L. Larson}
\affil{Department of Astronomy, The University of Texas at Austin, Austin, TX 78712, USA}

\author[0000-0001-7503-8482]{Casey Papovich}
\affil{Department of Physics and Astronomy, Texas A\&M University, College
Station, TX, 77843-4242 USA}
\affil{George P.\ and Cynthia Woods Mitchell Institute for
  Fundamental Physics and Astronomy, Texas A\&M University, College
  Station, TX, 77843-4242 USA}

\author[0000-0001-8940-6768]{Laura Pentericci}
\affil{INAF, Osservatorio Astronomico di Roma, via Frascati 33, 00078, Monteporzio Catone, Italy}

\author{Amber N. Straughn}
\affil{Astrophysics Science Division, Goddard Space Flight Center, Greenbelt, MD 20771, USA}

\author[0000-0003-2775-2002]{Yicheng Guo}
\affil{Department of Physics and Astronomy, University of Missouri, Columbia, MO 65211, USA}

\author[0000-0002-9226-5350]{Sangeeta Malhotra}
\affil{Astrophysics Science Division, Goddard Space Flight Center, Greenbelt, MD 20771, USA}
\affil{School of Earth \& Space Exploration, Arizona State University, Tempe, AZ 85287, USA}

\author[0000-0002-1501-454X]{James Rhoads}
\affil{Astrophysics Science Division, Goddard Space Flight Center, Greenbelt, MD 20771, USA}
\affil{School of Earth \& Space Exploration, Arizona State University, Tempe, AZ 85287, USA}

\author[0000-0002-8442-3128]{Mimi Song}
\affil{Department of Astronomy, University of Massachusetts, Amherst, MA 01002, USA}

\author[0000-0001-8514-7105]{Vithal Tilvi}
\affil{School of Earth \& Space Exploration, Arizona State University, Tempe, AZ 85287, USA}

\author[0000-0002-0784-1852]{Isak Wold}
\affil{Astrophysics Science Division, Goddard Space Flight Center, Greenbelt, MD 20771, USA}

\published{2020 November 27 in ApJ}

\begin{abstract}
Ly$\alpha$ emission from galaxies can be utilized to characterize the ionization state in the intergalactic medium (IGM).  We report our search for Ly$\alpha$ emission at $z>7$ using a comprehensive Keck/MOSFIRE near-infrared spectroscopic dataset, as part of the Texas Spectroscopic Search for Ly$\alpha$ Emission at the End of Reionization Survey. We analyze data from 10 nights of MOSFIRE observations which together target 72 high-$z$ candidate galaxies in the GOODS-N field, all with deep exposure times of 4.5--19 hr. Utilizing an improved automated emission-line search, we report 10 Ly$\alpha$ emission lines detected ($>$4$\sigma$) at $z>7$, significantly increasing the spectroscopically confirmed sample.  Our sample includes large equivalent-width (EW) Ly$\alpha$ emitters ($>$50\AA), and additional tentative Ly$\alpha$ emission lines detected at 3$\sigma$ -- 4$\sigma$ from five additional galaxies. We constrain the Ly$\alpha$ EW distribution at $z\sim7.6$, finding a significant drop from $z\lesssim6$, suggesting an increasing fraction of neutral hydrogen (\ion{H}{1}) in the IGM in this epoch. We estimate the Ly$\alpha$ transmission through the IGM ($=$EW$_{z\sim\text{7.6}}$/EW$_{z\sim\text{2--6}}$), and infer an IGM \ion{H}{1} fraction ($X_{\text{HI}}$) of $49^{+19}_{-19}\%$ at $z\sim7.6$, which is lower in modest tension ($>$1$\sigma$) with recent measurements at $z \sim$ 7.6.  The spatial distribution of the detected Ly$\alpha$ emitters implies the presence of a potential highly ionized region at $z\sim7.55$ which hosts four Ly$\alpha$ emitters within a $\sim$ 40 cMpc spatial separation. The prominence of this ionized region in our dataset could explain our lower inferred value of $X_{\text{HI}}$, though our analysis is also sensitive to the chosen reference Ly$\alpha$ EW distribution values and reionization models.
\end{abstract}
\keywords{early universe --- galaxies: distances and redshifts --- galaxies: evolution --- galaxies: formation --- galaxies: high-redshift --- intergalactic medium}

\section{Introduction}
Charting the timeline of reionization is a critical topic in observational cosmology. It also places a key constraint on the ionizing photon budget from galaxies that are thought to be dominating the supply of the required ionizing photons to make reionization happen \citep[e.g.,][]{Finkelstein2012b, Finkelstein2015a, Finkelstein2019b, Robertson2013a, Robertson2015a}. Although the Cosmic Microwave Background observations with Planck constrains the midpoint of reionization to be at $z\sim8$ \citep{Planck-Collaboration2016a}, and quasar observations suggest $z\sim6$ as the end point of reionization, a detailed study on how reionization proceeded over time is still lacking.  As Ly$\alpha$ emission visibility is sensitive to a changing amount of the neutral hydrogen (\ion{H}{1}) fraction in the IGM, it provides a way to derive the redshift evolution of the \ion{H}{1} fraction ($X_{\text{HI}}$) into the epoch of reionization \citep[e.g.,][]{Malhotra2004a, Stark2011a, Pentericci2011a, Dijkstra2014b, Konno2018a}.

Over the past decade, multiobject spectroscopic observations with large ground-based telescopes (e.g., Keck/DEIMOS, Keck/MOSFIRE, VLT/FORS2, VLT/KMOS, VLT/MUSE) have delivered a number of confirmed Ly$\alpha$ emitters (LAEs) at/around the end of reionization \citep[e.g.,][]{Finkelstein2013a, Schenker2014a, Tilvi2014a, Oesch2015a, Zitrin2015a, Song2016b, Herenz2017a, Hoag2017a, Laporte2017a, Stark2017a, Jung2018a, Jung2019a, Pentericci2018b, Mason2019a, Khusanova2020a}. Initial studies of the simple ``Ly$\alpha$ fraction'' ($=N_{\text{LAE}}/N_{\text{LBG}}$), where $N_{\text{LAE}}$ is the number of Ly$\alpha$-detected objects and $N_{\text{LBG}}$ is the number of high-$z$ candidate galaxies observed in spectroscopic observations, have found an apparent deficit of Ly$\alpha$ emission at $z > 6.5$ \citep[e.g.,][]{Fontana2010a, Pentericci2011a, Pentericci2018b}, implying an increasing \ion{H}{1} fraction in the IGM from $z\sim6$ $\rightarrow$ 7, although other Ly$\alpha$ systematics with galaxy evolutionary features need to be taken into account \citep[e.g.,][]{Yang2017a, Tang2019a, Trainor2019a, Du2020a}.

Using extensive Ly$\alpha$ spectroscopic data of $\gtrsim$ 60 Ly$\alpha$ detected galaxies over a wide-field area at $z \sim6 -7$, \cite{Pentericci2018b} suggest a smoother evolution of the IGM compared to previous studies, proposing that the IGM was not fully ionized by $z = 6$ \citep[see also ][]{Kulkarni2019a, Fuller2020a}. Furthermore, while \cite{Zheng2017a}, \cite{Castellano2018a}, and \cite{Tilvi2020a} report their observations of an ionized bubble via detection of multiple Ly$\alpha$ emitters at $z\sim7-8$, non/rare detections of Ly$\alpha$ in \cite{Hoag2019a} and \cite{Mason2019a} suggest a significantly neutral fraction in the IGM at $z\sim7.5$, with \cite{Hoag2019a} reporting a very high neutral fraction of 90\% at $z \sim$ 7.6.  Taken together, these results do not tell a coherent story. However, cosmic variance and the intrinsic inhomogeneity of the reionization process are likely playing at least a partial role. Reionization models predict that the spatial size of single ionized bubbles at $z \sim 7-8$ are $\sim$~10--20 cMpc or $\sim$~5$'$--10$'$ for $X_{\text{HI}}$=0.5 at $z\sim7$ \citep[e.g.,][]{Ocvirk2020a}, which is comparable to/larger than the field of view (FOV; $6'\times4'$) of MOSFIRE. Also,  previous observations of Ly$\alpha$ at this redshift may be too shallow \cite[e.g., half of the galaxies in][were observed for $\lesssim$3hr]{Hoag2019a}, which could result in lower detection rates.

Despite the recent accomplishments of Ly$\alpha$ spectroscopic studies as probes of reionization, they still require accounting for many forms of data incompleteness. First, the target selection solely depends on photometric redshift measurements, or the Lyman-break drop-out technique, which is less accurate at increasingly higher redshifts \citep[e.g.,][]{Pentericci2018b}. In addition, somewhat shallow observational depths limit Ly$\alpha$ detection, especially from faint sources. This is even more challenging at $z>7$, where observations shift into the NIR with its bright telluric emission, and the observable Ly$\alpha$ flux will be reduced even for low neutral fractions. 

Here we discuss the full results from our Texas Spectroscopic Search for Ly$\alpha$ Emission at the End of Reionization, which comprises 18 nights of spectroscopic observations with Keck/DEIMOS and MOSFIRE, targeting $\sim$200 galaxies at $z>5$.  In \cite{Jung2018a} we published the first result from our survey, introducing our methodology for constraining the evolution of the Ly$\alpha$ EW distribution accounting for all observational incompleteness effects (e.g., photometric redshift probability distribution function (PDF), UV-continuum luminosity, instrumental wavelength coverage, and observing depth). \cite{Jung2018a} constrained the Ly$\alpha$ EW distribution at $z\sim6.5$, finding a suggestion of a suppressed Ly$\alpha$ visibility and thus a sign of an increasing \ion{H}{1} fraction in the IGM.  The MOSFIRE portion of our dataset consists of 10 nights in the Great Observatories Origins Deep Survey North (GOODS-N) field in addition to 10hr of observation in the GOODS-S field published in \cite{Song2016b}. This MOSFIRE survey delivers near-infrared (NIR) Ly$\alpha$ spectroscopic observations for 84 galaxies with $t_{\text{exp}}\sim4.5$--19 hr, which results in the deepest and most comprehensive NIR Ly$\alpha$ spectroscopic survey at $z>7$.

In this study, we present our analysis on 10 nights of the MOSFIRE observations in the GOODS-N field, aiming to provide an observational constraint on the Ly$\alpha$ EW distribution at $z>7$. Section 2 describes the observational dataset, data reduction procedures, and target selection based on improved photometric redshift measurements \citep[S. L. Finkelstein et al. in preparation; also see][]{Finkelstein2013a, Finkelstein2015a}. In Section 3, we present the Ly$\alpha$ emission lines detected from our target galaxies, estimating their physical properties. Here we also implement an automated emission-line detection scheme to build a complete/unbiased emission-line catalog from spectroscopic data beyond visual inspection. Our measurement of the Ly$\alpha$ EW distribution at $z>7$ is shown in Section 4, and we discuss our constraints on reionization which include the \ion{H}{1} fraction and the ionization structure of the IGM in Section 5. Section 6 summarizes our findings. In this work, we assume the Planck cosmology \citep{Planck-Collaboration2016a} with $H_0$ = 67.8\,km\,s$^{-1}$\,Mpc$^{-1}$, $\Omega_{\text{M}}$ = 0.308, and $\Omega_{\Lambda}$ = 0.692.  The Hubble Space Telescope (HST) F435W, F606W, F775W, F814W, F850LP, F105W, F125W, F140W, and F160W bands are referred to as $B_{435}$, $V_{606}$, $i_{775}$, $I_{814}$, $z_{850}$, $Y_{105}$, $J_{125}$, $JH_{140}$ and $H_{160}$, respectively.  All magnitudes are given in the AB system \citep{Oke1983a}, and all errors presented in this paper represent 1$\sigma$ uncertainties (or central 68\% confidence ranges), unless stated otherwise.

\begin{figure*}[ht]
\centering
\includegraphics[width=1.05\columnwidth]{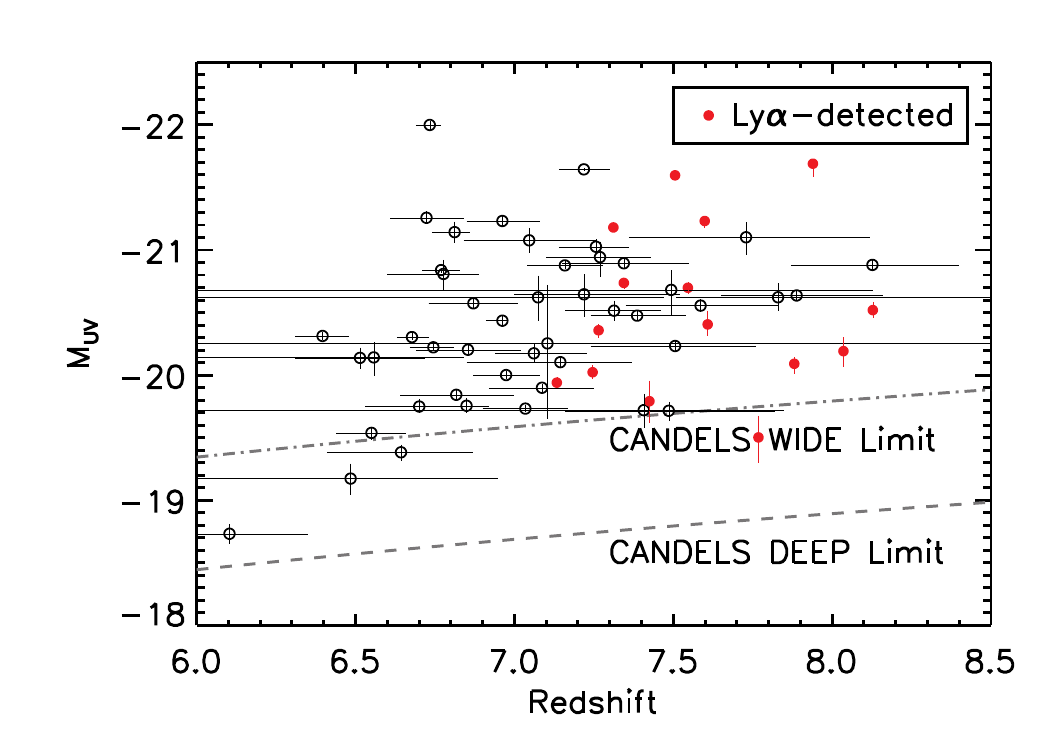}
\centering
\includegraphics[width=1.05\columnwidth]{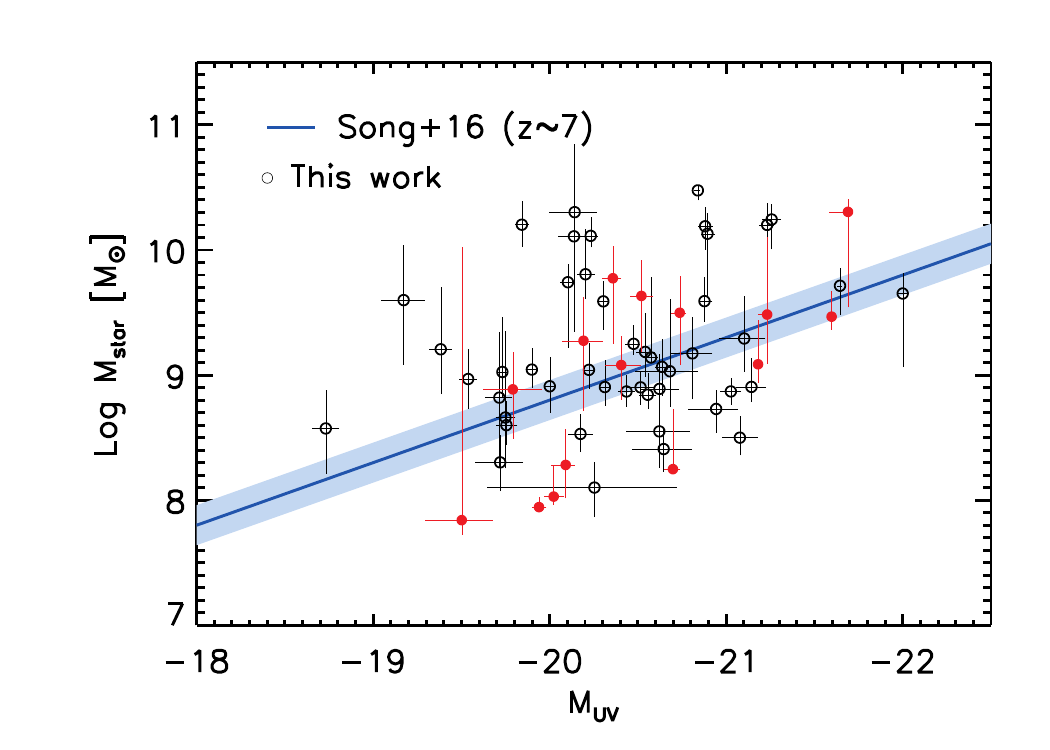}
\caption{(Left) $M_{\text{UV}}$ distribution of target galaxies in our MOSFIRE observations in the GOODS-N as a function of redshift.  Open circles display observed targets with photometric redshifts, and Ly$\alpha$ emitters are denoted as filled red circles with spectroscopic redshifts (Ly$\alpha$ emitters are further discussed in Section 3). The gray curves denote the limiting $M_{\text{UV}}$ in terms of the CANDELS/HST $J_{125}$ imaging depths: dashed and dotted-dashed lines are derived from the GOODS-N Deep and GOODS-N Wide fields, respectively.  The reason why we have no Ly$\alpha$ detection at $z<7$ is that the transmission curve of the MOSFIRE $Y$-band filter drops at $\sim$9800\AA\ (corresponding to Ly$\alpha$ at $z\sim7$).  (Right) $M_{\text{star}}$ -- $M_{\text{UV}}$ relation of the target galaxies. The blue solid line shows the fiducial $z\sim7$ relation of \cite{Song2016a} with the shaded area of its uncertainty.  Our spectroscopic sample is drawn fairly uniformly from this trend, showing no significant sample bias relative to the underlying galaxy population.}
\label{fig:MOSFIRE_muv}
\end{figure*}

\section{Data}
\subsection{Texas Spectroscopic Search for Ly$\alpha$ Emission at the End of Reionization}
Our spectroscopic data were obtained through a total of 18 nights of spectroscopic observations in the GOODS fields with Keck/DEIMOS \citep[PI: R. Livermore, published in][]{Jung2018a} and Keck/MOSFIRE (the majority awarded through the NASA/Keck allocation; PI: S. Finkelstein). The GOODS-S MOSFIRE observations were published in \cite{Song2016b}, and \cite{Jung2019a} published the deepest ($>$16hr) MOSFIRE dataset in GOODS-N.

\subsection{MOSFIRE $Y$-band Observations in GOODS-N}
In this study, we analyze the entire MOSFIRE dataset in GOODS-N, targeting 72 $z\gtrsim7$ galaxies over 10 nights of Keck/MOSFIRE observations with six mask designs from 2013 April to 2015 February. To cover Ly$\alpha$ over a redshift range of $z>7$, we used the $Y$-band filter which has a spectral resolution of $\sim3$\AA\ ($R=3500$). The slit width was chosen to be 0\farcs7 to match the typical seeing level at Maunakea.  During our observations, individual frames were taken with 180 s exposures with an ABAB dither pattern (+1\farcs25, -1\farcs25, +1\farcs25, -1\farcs25). The details of the observations are described in Table 1 of \cite{Jung2019a}. 

\subsection{Physical Properties of Target Galaxies}
Table \ref{tab:MOSFIREall} in the Appendix shows the list of the spectroscopic targets in our GOODS-N MOSFIRE observations. The target selection was based on the photometric redshift catalog of \cite{Finkelstein2013a, Finkelstein2015a}, utilizing the HST/CANDELS photometric catalog \citep{Grogin2011a, Koekemoer2011a}. Slitmask configurations were designed by MAGMA,\footnote{https://www2.keck.hawaii.edu/inst/mosfire/magma.html} maximizing the Ly$\alpha$ detection probability based on the galaxy brightness and the photometric redshift probability within the instrumental wavelength coverages. Although the MOSFIRE $Y$-band coverage for Ly$\alpha$ is limited at $z>7$, we include $z \gtrsim 6.5$ galaxies in the target selection, accounting for the photometric redshift uncertainties.  At the time of target selection, the redshift information was based on the previous version of the photometric redshift catalog in \cite{Finkelstein2015a}.  Recently, S. L. Finkelstein et al. (in preparation) has updated the photometric redshift measurements with updated CANDELS photometry including deep $I_{814}$ and Spitzer/IRAC data where they performed deblending of the low-resolution IRAC images with the HST images as priors. We use the updated photometric redshift catalog of S.L. Finkelstein et al. (in preparation) for our analysis, and 10 observed galaxies are now likely to be low-$z$ objects in the updated catalog and are excluded from the analysis for the remainder of this study.\footnote{The low-z targets are listed at the bottom in Table \ref{tab:MOSFIREall}}

To understand the properties of our observed sources, we perform spectral energy distribution (SED) fitting with the \cite{Bruzual2003a} stellar population model. We utilize \emph{HST}/ACS ($B_{435}$, $V_{606}$, $i_{775}$, $I_{814}$ and $z_{850}$) and WFC3 ($Y_{105}$, $J_{125}$, $JH_{140}$ and $H_{160}$) broadband photometry and Spitzer/IRAC 3.6$\mu$m and 4.5$\mu$m band fluxes. We assume a \cite{Salpeter1955a} initial mass function with a stellar mass range of 0.1--100$M_{\odot}$, and a metallicity range of 0.005--1.0$Z_{\odot}$. We adopt a range of exponential models of star formation histories with various exponentially varying timescales ($\tau=$ 10 Myr, 100 Myr, 1 Gyr, 10 Gyr, 100 Gyr, -300Myr, -1 Gyr. -10 Gyr). We model galaxy spectra using the \cite{Calzetti2001a} dust attenuation curve with $E(B-V)$ values ranging from 0 to 0.8, and nebular emission lines are added as described in \cite{Salmon2015a}, which adopts the \cite{Inoue2011a} emission-line ratios. The intergalactic medium attenuation is considered based on \cite{Madau1995a}. The best-fit models have been obtained minimizing $\chi^2$, and the uncertainties of physical parameters are estimated by performing SED fitting with 1000 Monte Carlo (MC) realizations of the perturbed photometric fluxes for individual galaxies. For the Ly$\alpha$-detected objects, we fit the model SEDs to the observed fluxes after subtracting the Ly$\alpha$ contributions in the continuum fluxes.  UV magnitudes ($M_{\text{UV}}$) of galaxies are estimated from the averaged fluxes over a 1450--1550\AA\ bandpass from the best-fit models. The best-fit model SEDs of our target galaxies are displayed in Figure \ref{fig:galaxy_seds1} and \ref{fig:galaxy_seds2} in the Appendix with the observed photometry. 

The left panel in Figure \ref{fig:MOSFIRE_muv} shows the $M_{\text{UV}}$ distribution of our GOODS-N MOSFIRE targets as a function of redshift. The black circles show the entire sample, and the red circles denote Ly$\alpha$-detected objects. As shown in the figure, our targets are randomly distributed over a wide range of $-19 \lesssim M_{\text{UV}} \lesssim -22$ with fewer faint objects at increasing redshift. This is somewhat natural due to the limiting observational depths in the continuum observations at higher redshifts. The reason why we have no Ly$\alpha$ detection at $z<7$ is that the transmission curve of the MOSFIRE $Y$-band filter drops at $\sim$~9800\AA\ (corresponding to Ly$\alpha$ at $z\sim7$). In the right panel, we display the $M_{\text{star}}$--$M_{\text{UV}}$ relation of our targets. Our galaxies are scattered out to broad regions in the relation, but consistent to the fiducial $z\sim7$ measurement of \cite{Song2016a}. Overall, our target selection does not exhibit a significant selection bias, representing the typical high-$z$ galaxy population at that redshift.

\subsection{Data Reduction and Flux Calibration}
\begin{figure}[t]
\centering
\includegraphics[width=1\columnwidth]{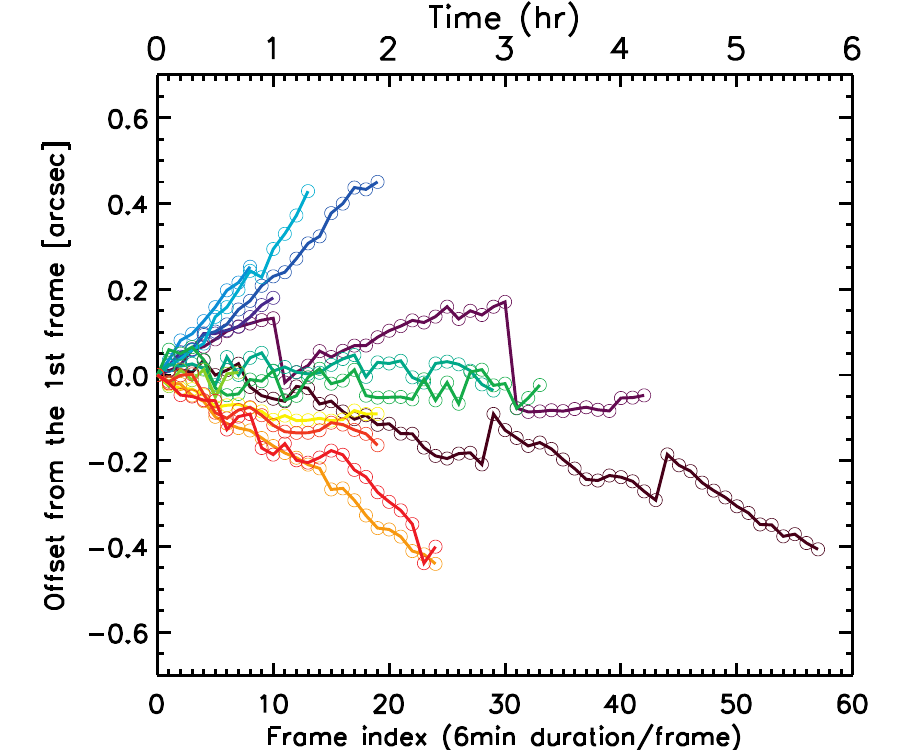}
\caption{Offsets in the spatial direction from the first frame as a function of time,  showing that objects drift in MOSFIRE slits along the spatial direction. This was corrected for during our data reduction, as described in Section  2.4.  Different colors represent different nights of observations.}
\label{fig:slit_drift}
\end{figure}

We reduced the raw data using the most recent version of the public MOSFIRE data reduction pipeline (DRP)\footnote{http://keck-datareductionpipelines.github.io/MosfireDRP/}. The DRP provides a sky-subtracted, flat-fielded, and rectified slit spectrum per object with a wavelength solution based on telluric sky emission lines. In the reduced two-dimensional (2D) spectra, the spectral dispersion is 1.09\AA\ pixel$^{-1}$, and the spatial resolution is 0\farcs18 pixel$^{-1}$. However, a noticeable slit drift in the spatial direction has been reported in previous MOSFIRE observations \citep[e.g.,][]{Kriek2015a, Song2016b, Jung2019a}, and we also detected slit drifts of up to $\sim$pixel hr$^{-1}$. To correct for this slit drift, we reduced each adjacent pair of science frames separately, generating reduced 2D spectra for 360 s of integration time. We estimated the relative slit drift by tracing the positions of slit continuum sources (either stars or bright filler galaxies) in the spatial direction on the DRP-produced 2D spectra. Figure \ref{fig:slit_drift} shows the measured offsets in the spatial direction from the first frame as a function of time. Colors represent individual nights. The measured drifts are up to $\sim$~$0\farcs1$~hr$^{-1}$. As this drift was a known issue at the time of our MOSFIRE observations, we aligned the telescope repeatedly in every 1--2 hr during some of the observations, shown as the breaking points in the plot.

The measured slit drift was corrected later when combining the individual DRP outputs. Running the DRP with a pair of frames makes it difficult to clean cosmic rays (CR) or bad pixels, thus we take sigma-clipped means in the combinations step in order to reject the bad pixels and CRs. To achieve an optimal signal-to-noise ratio (S/N), we weight the DRP outputs with the best-fit Gaussian peak fluxes of the continuum sources, which reflect observing conditions (e.g., seeing and transparency). 

We performed long-slit observations of spectrophotometric standard stars for flux calibration and telluric absorption correction using \cite{Kurucz1993a} model stellar spectra. To obtain the response curve as a function of wavelength, we divided the model stellar spectra, which were scaled to match with the known photometric magnitudes of the standard stars, with the long-slit stellar spectra. We also corrected slit losses via flux calibration, considering the seeing condition of each night of observations.  

As our science masks were observed in somewhat different observing conditions than the long-slit standard stars due to changing atmospheric conditions and airmass, we used continuum sources (stars) in our science masks to check for any residual flux calibration offset.  We first applied the flux calibration from the spectrophotometric standard to our slitmask stars, then integrated these spectra through the {\it HST}/WFC3 $Y_{105}$ bandpass, and then compared these values to the $Y$-band magnitudes are taken from the updated photometric catalog of S. L. Finkelstein et al. (in preparation) based on the HST/CANDELS photometric data.  We derived a residual normalization correction as the $Y$-band flux ratio between the known flux of these stars, and those derived from our MOSFIRE spectrum, such that after this correction was applied, they had the same $Y$-band magnitude.  This residual normalization was up to $\sim$30-50\%. As this correction can result in additional systematic errors in the flux calibration, it is recommended to have multiple continuum sources in science masks for future observations.  

Each night of observations was calibrated individually, and some of the science masks were observed for multiple nights. We combined data from these masks after flux calibration, weighted with the best-fit Gaussian peak fluxes of the slit continuum sources. Also, we observed 49 galaxies in multiple masks; thus, these were combined with a weight factor based on the median-noise levels to generate a single fully reduced, all-mask-combined, and flux-calibrated 2D spectrum per object. The one-dimensional (1D) spectra of the sources were obtained via an optimal extraction \citep{Horne1986a} with a 1\farcs4 spatial aperture, twice the typical seeing level from our observations. For the optimal extraction, we built a spatial weight profile from the stellar profile so that the pixels near the peak of the stellar profile were maximally weighted. 

\section{Results}
\subsection{Ly$\alpha$ Detections from an Automated Line Search}
Although Ly$\alpha$ has been proven to be a useful method for confirming the redshifts of high-$z$ candidate galaxies, it becomes very challenging to detect into the epoch of reionization as it is sensitive to an increasing amount of neutral hydrogen in the IGM, and also becomes fainter as it is coming from more distant objects. Due to such hurdles, there have been only 10 Ly$\alpha$-emitting galaxies so far detected at $z>7.5$ \citep{Finkelstein2013a, Oesch2015a, Zitrin2015a, Song2016b, Laporte2017a, Hoag2017a, Hashimoto2018a, Jung2019a, Tilvi2020a}.

Another technical challenge of Ly$\alpha$ spectroscopic follow-up observations is in the search for faint emission-line features from obtained spectra, as it is difficult to distinguish them from noise peaks with human eyes. To perform a thorough scan on observed spectra, an automated approach has been adopted in previous studies \citep[e.g.,][]{Jung2018a, Larson2018a, Pentericci2018b, Hoag2019a}, which can play a supplemental role to visual inspection, capturing missing plausible features. 

\begin{figure}[t]
\centering
\includegraphics[width=1\columnwidth]{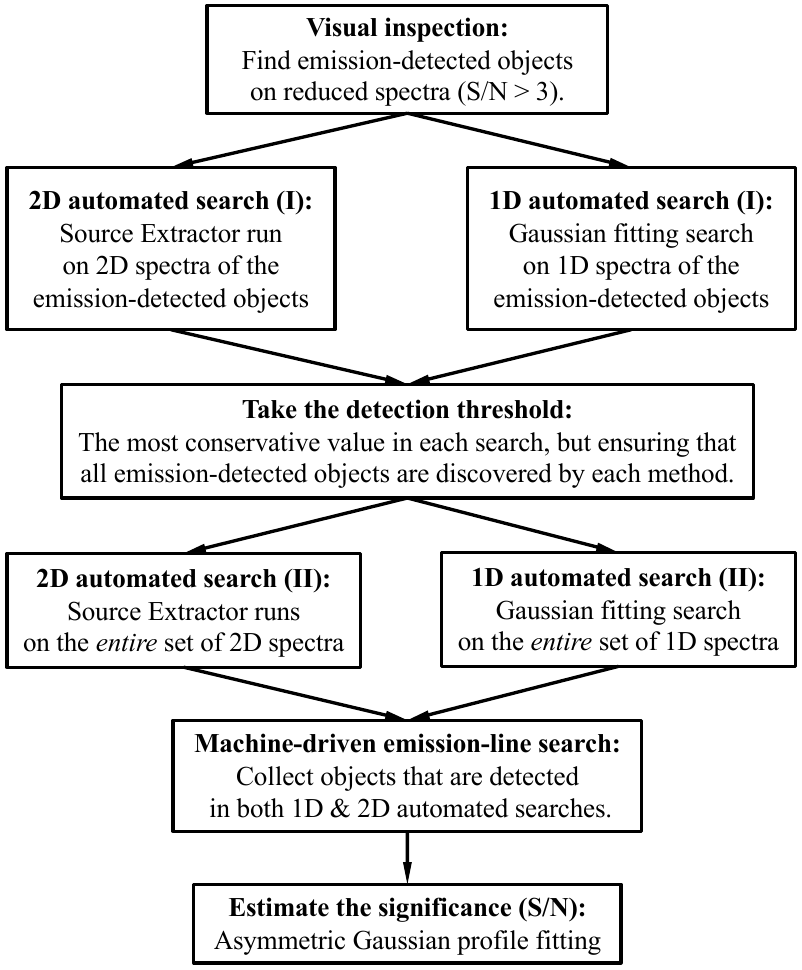}
\caption{Graphical representation of our automated emission-line search procedures. The 1D automated search algorithm is adopted from \cite{Larson2018a}.}
\label{fig:auto_search}
\end{figure}

In this work, we attempt to perform an improved automated search using the Source Extractor software \citep[SExtractor;][]{Bertin1996a} on 2D spectra as well as Gaussian line fitting on 1D spectra \citep[e.g.,][]{Larson2018a}. Figure \ref{fig:auto_search} summarizes the entire procedure of our automated emission-line search. First we performed several iterations of visual inspection to search for any significant emission-line features, and estimated their detection levels as the S/N values of Ly$\alpha$ emission fluxes. To estimate emission-line properties, we performed asymmetric Gaussian fitting on reduced 1D spectra with the IDL \texttt{MPFIT} package \citep{Markwardt2009a}.  The asymmetric Gaussian function is defined as
\begin{equation}
f(\lambda) = f_0\times \begin{cases} 
	\exp\Big{(}-\frac{1}{2}\frac{(\lambda-\lambda_0)^2}{\sigma_{\text{blue}}^2}\Big{)}& \text{for}\ \lambda \leq \lambda_0,\\
	\exp\Big{(}-\frac{1}{2}\frac{(\lambda-\lambda_0)^2}{\sigma_{\text{red}}^2}\Big{)}& \text{for}\ \lambda > \lambda_0,
	\end{cases}
\end{equation}
where $f_0$ is the peak value of the profile, $\lambda_0$ is the peak wavelength, $\sigma_{\text{blue}}$ and $\sigma_{\text{red}}$ are the blue- and red-side widths of the profile. In the fitting procedure, we have a zero continuum flux prior with initial guesses of $f_0=0.5\times10^{-18}$ erg s$^{-1}$ cm$^{-2}$,  $\lambda_0$ is at the peak flux wavelength in a 1D spectrum, and $\sigma_\text{blue, red} = 3$\AA. We adjust the wavelength range that is included in the fit for achieving the maximum S/N of the emission line by reducing nearby contaminations while still capturing the entire emission-line contribution.
The associated errors of the physical quantities were derived from MC simulations by fluctuating the 1D spectra within their noise levels. From this visual inspection, we detect 13 emission-line features above a 3$\sigma$ level: 8 with S/N$>$4 and 5 with 3$<$S/N$<$4. 

Then, to catch any emission features missed by the previous visual inspection, we applied the automated line search scheme of \cite{Larson2018a} to the reduced 1D spectra and also ran Source Extractor on the reduced 2D flux and noise spectra. In the Source Extractor runs, we adjusted parameters to optimally search for point sources in 2D spectra with sizes comparable to the seeing value, typically $\sim$~0$\farcs7$ or 4 pixels in the spatial direction of the 2D spectra. In both 1D and 2D searches, we disregarded sky emission-line regions to avoid spurious emission features from sky residuals. However, the choice of a detection threshold in the automated searches is still arbitrary. A lower cut provides many more emission lines, which still include numerous false emission lines of, for example, uncleaned CRs, noise spikes, or contamination from nearby sources, while increasing the detection threshold would lose actual emission lines.  Thus, we elected to place the highest detection thresholds where the automated searches still capture all of the 13 significant emission lines from our visual inspection, which is $\gtrsim$3$\sigma$ in the 1D search and $\gtrsim$4$\sigma$ in the Source Extractor runs.\footnote{Flux error from Source Extractor's automated aperture photometry.}

With these detection thresholds, we examined the results of the 1D and 2D automated searches and found 29 emission features that were simultaneously detected by both 1D and 2D searches. This includes all previously detected emission lines from the visual inspection except for one tentative $\sim$3.5$\sigma$ detection from z8\_GND\_41470, as it was found very close to a sky emission line. Thus, applying the automated scans on 1D and 2D spectra found 17 additional plausible emission lines. Our improved automated search allows us to perform a machine-driven consistent emission-line search, where all plausible emission lines passed both automated searches with the same detection threshold as those from the visual inspection. Lastly, we measured S/N values for all plausible emission lines by performing asymmetric Gaussian fitting, which finds that 22 of these 29 lines have S/N$>$3; we discarded the 7 lower S/N sources. We visually inspected these 22 S/N $>$ 3 emission lines and found that five appeared to be sky residuals, and one appears spurious as it does not have the accompanying negative peaks expected for real sources due to our dithering sequence. Thus, our sample consists of 16 emission lines at $>$3$\sigma$ significance from this automated scanning, in addition to z8\_GND\_41470 (found visually). This automated search added three emission lines at a 4$\sigma$ level and one detection at a 3$\sigma$--4$\sigma$ level, which were not detected in visual inspection. This results in 17 significant emission lines: 11 with S/N$>$4 and 6 with 3$<$S/N$<$4.

\begin{deluxetable*}{ccccccccccc}
\setlength{\tabcolsep}{0.03in}
\tablewidth{0pc}
\tabletypesize{\footnotesize}
\tablecaption{Summary of emission-line Properties\tablenotemark{a} \label{tab:MOSFIRE_LAE_spectra}} 
\tablehead{
\colhead{ID} & \colhead{$F_{\text{Ly}\alpha}$}   & \colhead{S/N}   & \colhead{EW$_{\text{Ly}\alpha}$\tablenotemark{b}} & \colhead{$z_{\text{Ly}\alpha}$} & \colhead{$L_{\text{Ly}\alpha}$} & \colhead{\ion{H}{2} Radii} & \colhead{$M_{\text{UV}}$} & \colhead{FWHM$_{\text{red}}$\tablenotemark{c}} & \colhead{Log$(\sigma_{\text{red}}/\sigma_{\text{blue}})$} & {$\chi^2_{\text{Reduced}}$} \\
	\colhead{}  & \colhead{(10$^{-17}$ erg s$^{-1}$ cm$^{-2}$)} & \colhead{} & \colhead{(\AA)} & \colhead{} &\colhead{(10$^{43}$ erg s$^{-1}$)}  & \colhead{(pMpc)}&\colhead{}  &\colhead{(km s$^{-1}$)} &\colhead{} &\colhead{$z_{\text{Ly$\alpha$}}(z_{\text{[\ion{O}{2}]}})$} \\
	\colhead{(1)}  & \colhead{(2)} & \colhead{(3)} & \colhead{(4)} & \colhead{(5)} & \colhead{(6)} & \colhead{(7)} & \colhead{(8)} & \colhead{(9)} & \colhead{(10)}& \colhead{(11)}
}
\startdata
{z7\_GND\_44088} & {1.27$\pm$0.25} & {5.2} 				& {87.6$^{+23.8}_{-21.2}$} &{7.1335$\pm$0.0028} &{0.79$\pm$0.16} &{0.88} &{-19.9} &{277$^{+69}_{-91}$}&{0.77$^{+3.37}_{-1.02}$}  &{1.2 (7.2)}\\  
{z8\_GND\_22233} & {1.36$\pm$0.19} & {7.1} 				& {54.5$^{+15.0}_{-12.1}$} &{7.3444$\pm$0.0020} &{0.91$\pm$0.13} &{0.92} &{-20.7} &{264$^{+75}_{-18}$}&{0.99$^{+2.99}_{-0.89}$}    &{2.6 (25.8)}\\  
{z7\_GND\_18626} & {0.26$\pm$0.06} & {4.6} 				& {26.8$^{+14.9}_{-9.8}$} &{7.4249$\pm$0.0013} &{0.18$\pm$0.04} &{0.52} &{-19.8} &{93$^{+80}_{-71}$}&{3.54$^{+0.97}_{-3.12}$}    &{0.9 (2.8)}\\  
{z7\_GND\_42912\tablenotemark{d}} & {1.46$\pm$0.13} & {10.8} & {33.2$^{+4.3}_{-4.0}$} &{7.5056$\pm$0.0007} &{1.02$\pm$0.09} &{0.96} &{-21.6} &{266$^{+57}_{-61}$}&{0.47$^{+0.19}_{-0.21}$}    &{1.5 (23.5)}\\  
{z7\_GND\_6330} & {0.41$\pm$0.07} & {6.1} 				& {15.9$^{+4.4}_{-3.7}$} &{7.5460$\pm$0.0006} &{0.29$\pm$0.05} &{0.62} &{-20.7} &{$<$88}&{0.43$^{+2.33}_{-0.50}$}    &{1.3 (16.9)}\\  
{z7\_GND\_16863\tablenotemark{e}} & {1.89$\pm$0.18} & {10.8} & {61.3$^{+14.4}_{-11.4}$} &{7.5989$\pm$0.0011} &{1.36$\pm$0.13} &{1.07} &{-21.2} &{411$^{+6}_{-54}$}&{1.42$^{+3.07}_{-0.80}$}    &{1.2 (22.4)}\\  
{z7\_GND\_34204} & {4.51$\pm$0.57} & {7.9} 				& {279.7$^{+80.4}_{-62.5}$} &{7.6082$\pm$0.0030} &{3.26$\pm$0.41} &{1.44} &{-20.4} &{365$^{+141}_{-88}$}&{0.27$^{+0.60}_{-0.31}$}    &{1.2 (5.2)}\\  
{z7\_GND\_7376} & {0.26$\pm$0.06} & {4.1} 				& {32.5$^{+23.0}_{-13.0}$} &{7.7681$\pm$0.0024} &{0.20$\pm$0.05} &{0.54} &{-19.5} &{147$^{+69}_{-99}$}&{0.35$^{+3.48}_{-0.89}$}    &{0.8 (1.9)}\\  
{z7\_GND\_39781} & {1.60$\pm$0.35} & {4.5} 				& {123.9$^{+37.4}_{-32.9}$} &{7.8809$\pm$0.0018} &{1.25$\pm$0.27} &{1.04} &{-20.2} &{$<$85}&{-0.17$^{+1.10}_{-3.25}$}    &{0.6 (11.0)}\\
{z7\_GND\_10402\tablenotemark{f}} & {0.32$\pm$0.08} & {4.0} & {6.7$^{+2.7}_{-2.2}$} &{7.9395$\pm$0.0023} &{0.26$\pm$0.06} &{0.59} &{-21.7} &{107$^{+44}_{-92}$}&{0.94$^{+5.40}_{-2.30}$}    &{0.1 (97.0)}\\  
\hline
{z7\_GND\_6451} & {0.68$\pm$0.21} & {3.2} 				& {43.2$^{+17.2}_{-15.1}$} &{7.2462$\pm$0.0045} &{0.44$\pm$0.14} &{0.72} &{-20.0} &{93$^{+65}_{-70}$}&{0.20$^{+0.97}_{-1.22}$}    &{0.6 (9.6)}\\  
{z7\_GND\_45190} & {0.31$\pm$0.09} & {3.4} 				& {22.9$^{+13.7}_{-9.3}$} &{7.2650$\pm$0.0019} &{0.20$\pm$0.06} &{0.55} &{-20.4} &{$<$91}&{-0.31$^{+3.91}_{-2.75}$}    &{1.2 (8.6)}\\  
{z8\_GND\_41470} & {0.93$\pm$0.26} & {3.5} 				& {25.9$^{+9.5}_{-8.4}$} &{7.3115$\pm$0.0028} &{0.61$\pm$0.17} &{0.81} &{-21.2} &{$<$90}&{-1.21$^{+1.69}_{-3.54}$}    &{5.9 (36.7)}\\  
{z8\_GND\_41247} & {1.70$\pm$0.44} & {3.9} 				& {164.2$^{+85.8}_{-60.5}$} &{8.0356$\pm$0.0015} &{1.39$\pm$0.36} &{1.07} &{-20.2} &{$<$83}&{-0.42$^{+1.16}_{-3.36}$}    &{0.1 (2.1)}\\  
{z7\_GND\_7157} & {0.26$\pm$0.08} & {3.4} 				& {21.2$^{+9.9}_{-8.0}$} &{8.1280$\pm$0.0016} &{0.22$\pm$0.07} &{0.56} &{-20.5} &{161$^{+59}_{-127}$}&{1.28$^{+3.70}_{-1.55}$}    &{2.0 (3.4)}\\  
\enddata
\tablecomments{Col. (1) Object ID, (2) Ly$\alpha$ emission-line flux, (3) signal-to-noise ratio, (4) equivalent width of the Ly$\alpha$ emission line, (5) spectroscopic redshift based on the Ly$\alpha$ emission line, (6) Ly$\alpha$ emission luminosity, (7) radii of ionized \ion{H}{2} bubbles around LAEs, based on the relation between Ly$\alpha$ luminosities and the bubble sizes from the \cite{Yajima2018a} model (see more discussion in Section 5), (8) galaxy UV magnitude, estimated from the averaged flux over a 1450--1550\AA\ bandpass from the best-fit galaxy SED model, (9) velocity FWHM, inferred from the red side of the emission line, corrected for the instrumental broadening, (10) asymmetry of the Ly$\alpha$ emission-line profile, where $\sigma_{\text{blue}}$ and $\sigma_{\text{red}}$ represent the blue- and red-side line widths, and (11) reduced $\chi^2$ values from the best-fit SED models of $z_{\text{Ly$\alpha$}}$ ($z_{\text{[\ion{O}{2}]}}$).\\
$^{a}$\footnotesize Five emission lines with S/N$<$4, listed at the bottom, are not included in the remainder of our analysis in Section 4 and 5.\\
$^{b}$Listed uncertainties account for the UV-continuum measurement errors from SED fitting.\\
$^{c}$In case that the measured values are smaller than the instrumental broadening, we provide the instrumental broadening as an upper limit.\\
$^{d}$Known as z8\_GND\_5296 in \cite{Finkelstein2013a} and updated in \cite{Jung2019a}. The source was observed in \cite{Tilvi2016a} and \cite{Hutchison2019a} as well.\\
$^{e}$Ly$\alpha$ emission reported in \cite{Jung2019a}.\\
$^{f}$Keck/DEIMOS observations of \cite{Jung2018a} displayed an emission-line feature at $z_{\text{Ly$\alpha$}}=6.70$, but it turned out to be contaminating \ion{O}{3} emission from a nearby $z=0.87$ object (noted by Lennox Cowie).}
\end{deluxetable*}

\subsection{Low-$z$ Interpretations}
To further explore whether these lines are indeed Ly$\alpha$, we checked the possibility of other, lower-redshift, solutions. First, we checked multiple-emission-line objects with the other emission lines (e.g., [\ion{O}{3}] $\lambda\lambda$4959, 5007, H$\beta$, [\ion{N}{2}] $\lambda\lambda$6548, 6584, H$\alpha$) as the MOSFIRE spectral coverage would allow us to detect multiple lines in these low-$z$ objects. However, none of our objects show multiple emission lines from our emission-line search, which rules out their possibilities of being those emission lines listed except for either Ly$\alpha$ or  [\ion{O}{2}] $\lambda\lambda$3727, 3729.

Specifically, the strong continuum break with an emission line between the optical and NIR photometric bands indicates that the emission is either Ly$\alpha$ with the Lyman break or [\ion{O}{2}] $\lambda\lambda$3727, 3729 with the Balmer break.  In the case of the [\ion{O}{2}] doublet, a peak separation of the doublet based on the wavelength in vacuum is 2.783\AA\ from the atomic line List (from www.pa.uky.edu), and it would be $\sim$7 -- 8\AA\ at $z\sim1.6$--$2.0$.  Given the spectral resolution of Keck/MOSFIRE ($R=3500$ or $\sim$3\AA), the doublet should be resolved if the emission line was indeed [\ion{O}{2}].  Although no object displays the doublet emission lines, it is still possible that these might be [\ion{O}{2}] emission. Due to many sky emission lines in the NIR and the somewhat low S/Ns of the detected lines, only one of the doublet could be detected while the other peak is either too faint to be detected or located in sky emission regions. Therefore, we performed SED fitting for all emission-line-detected galaxies using broadband photometry, fixing the redshift to both the high-$z$ (Ly$\alpha$) and low-$z$ ([\ion{O}{2}]) solutions. We compared $\chi^2_{\text{reduced}}$ values between the high- and low-$z$ emission-line solutions (listed in Table \ref{tab:MOSFIRE_LAE_spectra}) and removed two galaxies (z7\_GND\_43678 and z7\_GND\_36688) that do not clearly prefer high-$z$ solutions, as they exhibit significant fluxes in HST photometric bands which are blueward of the emission-line wavelength, which is not expected were the line Ly$\alpha$.  Specifically, z7\_GND\_43678 displays a very poor SED fit ($\chi^2_{\text{reduced}}>$10) for the high-$z$ case, and its $z_{\text{Ly$\alpha$}}=7.69$ is greatly mismatched to its tightly constrained $z_{\text{phot}}=6.40^{+0.08}_{-0.09}$ (based on the detections in $I_{814}$ and $z_{850}$).  We find similar results for z7\_GND\_36688 ($\chi^2_{\text{reduced}}>$5 for the high-$z$ solution). Additionally, the emission line from z7\_GND\_36688 is relatively less certain to be a real emission line with its low S/N = 3.6 while that from z7\_GND\_43678 appears more convincing (S/N = 5.4).

\begin{figure*}[t]
\centering
\includegraphics[width=0.84\paperwidth]{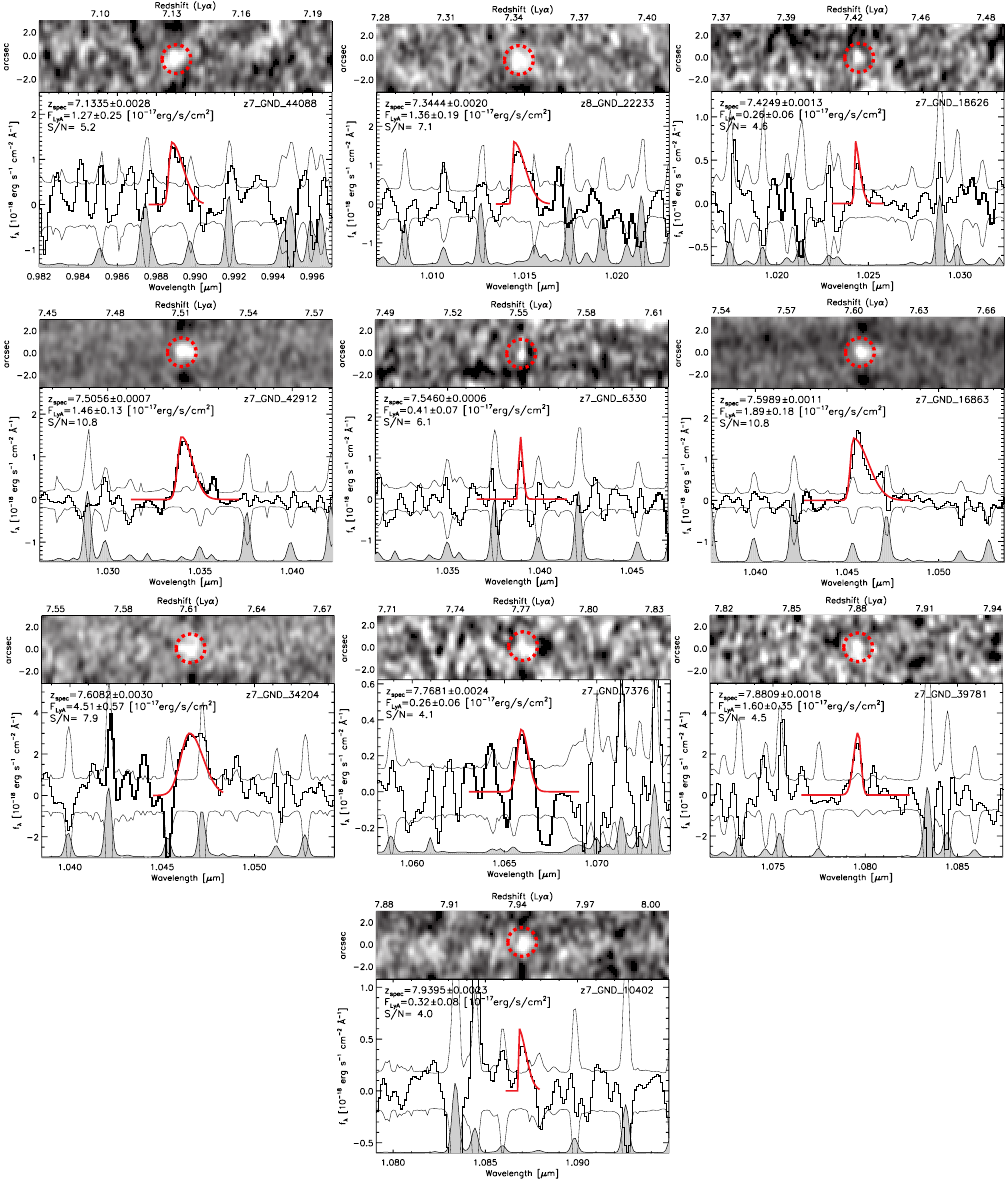}
\caption{\small One- (bottom) and two-dimensional (top) spectra of our detected Ly$\alpha$ emission lines (S/N $>4\sigma$). In the 1D spectra, the black histogram is the all-mask-combined flux, smoothed by the instrumental resolution of $\sim3$\AA. The thin black curves represent the $1\sigma$ noise level, and the normalized sky emission is plotted at the bottom as a gray filled curve. Red curves show the best-fit asymmetric Gaussian curves. The displayed 2D spectra are the S/N maps that were obtained by dividing the sky-subtracted spectra by the noise spectra. The 2D spectra are smoothed by the instrumental resolution (x direction) and the seeing of the observations (y direction), and the dotted red circles denote detected emission lines.}
\label{fig:MOSLAEspectra}
\end{figure*}

\begin{figure*}[t]
\centering
\includegraphics[width=0.84\paperwidth]{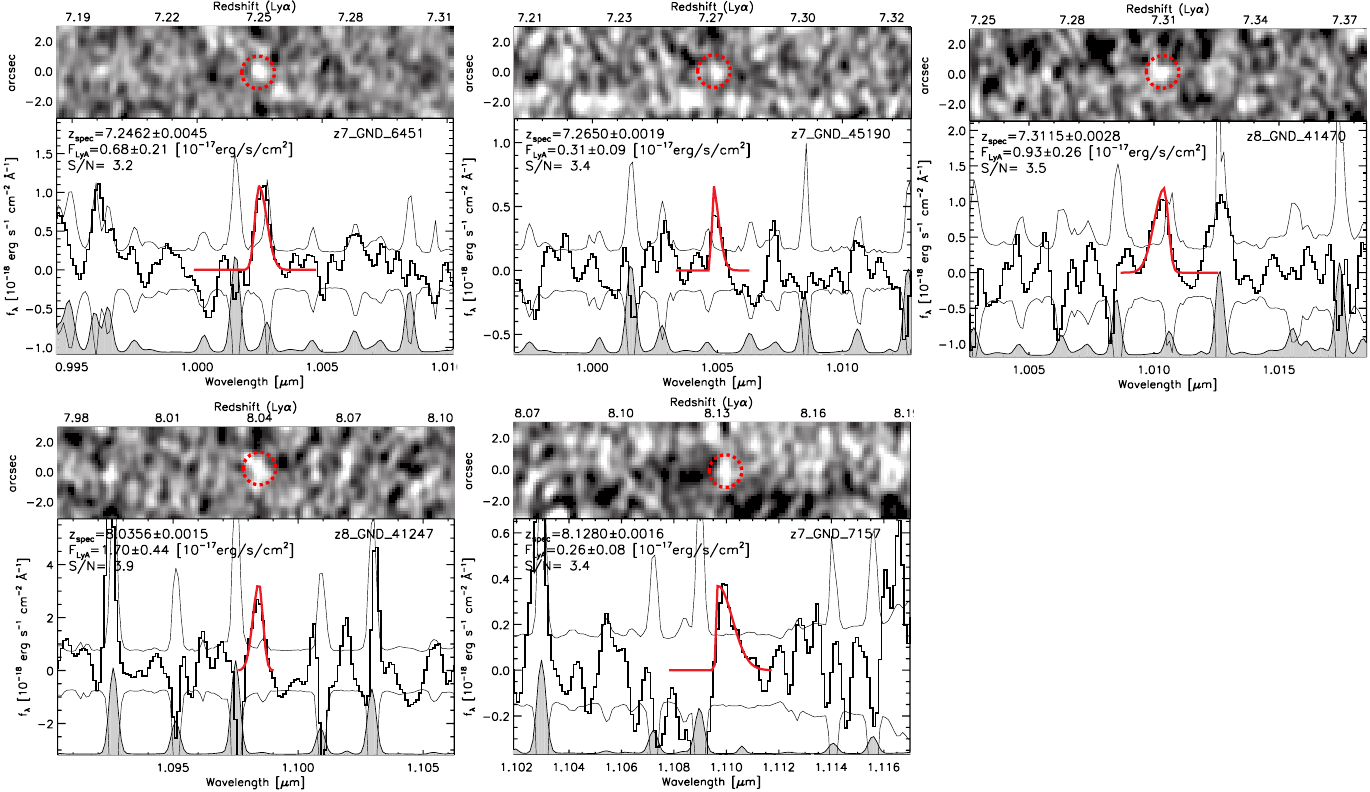}
\caption{Same as Figure \ref{fig:MOSLAEspectra} but with S/N = 3$\sigma$--4$\sigma$.}
\label{fig:MOSLAEspectra34}
\end{figure*}

We attempted to identify these emission lines among the possible low-$z$ interpretations. However, a robust identification is difficult as we detect neither a double peak indicating [\ion{O}{2}] emission nor multiple emission lines (indicating, e.g., H$\alpha$ with [\ion{N}{2}]) from these galaxies.  Their SED fits are also poor at their [\ion{O}{2}] and H$\alpha$ emission redshifts, additional perplexities that remain on these emission lines.  Thus, while we cannot identify which emission lines these are, we do not include either source in our list of Ly$\alpha$-emitting galaxies.  Finally, we have 15 objects which show Ly$\alpha$ emission ($>$3$\sigma$) after this low-$z$ check: 10 with S/N$>$4 and 5 with 3$<$S/N$<$4. The 1D and 2D spectra of our Ly$\alpha$ emitters are displayed in Figure \ref{fig:MOSLAEspectra} and \ref{fig:MOSLAEspectra34}.

\subsection{False Detection Check in Automated Line Search}
With the true spectra, our automated search provided 13 emission lines at S/N$>$4, and 11 of them were found to be actual emission lines with only one spurious source without negative counterparts and one another sky residual.  To further explore the rate of spurious contamination in our emission-line selection process, we performed our automated search with the same detection criteria on negative versions of the 1D and 2D spectra.  Across the negative versions of our observed galaxy spectra, we found eight S/N$>$4 emission lines.  However, only one appears to represent a truly spurious detection.  Two are sky residuals, and the other five are negative peaks caused by nearby contaminating objects, all of which would be flagged in our visual inspection on the true galaxy spectra.  Thus, we conclude that our automated line search successfully delivers real emission lines with minimal spurious detection, although we still perform supplemental cleaning of sky residuals and spurious sources by visual inspection.

In addition, our 10 S/N$>$4 Ly$\alpha$ emission lines in Table \ref{tab:MOSFIRE_LAE_spectra} and Figure \ref{fig:MOSLAEspectra} are not likely spurious but securely detected Ly$\alpha$ as their 2D spectra display negative peaks on the top and the bottom sides, which emerge from the dithering pattern of the observations and would not be present for a spurious signal. We classify the five emission lines below 4$\sigma$ as tentative, needing further verification with deeper observations, and we do not include them in the remainder of our analysis for constraining the Ly$\alpha$ EW distribution. It is worth mentioning that if we perform the same analysis with all 15 emission lines, it does not significantly change our results and major conclusions.

\subsection{Ly$\alpha$ Emission Properties}
We derive the physical quantities of the detected emission lines by performing asymmetric Gaussian fitting on reduced 1D spectra, as described in the previous section. The derived line properties are summarized in Table \ref{tab:MOSFIRE_LAE_spectra}.  

\begin{figure*}[ht!]
\centering
\includegraphics[width=0.62\paperwidth]{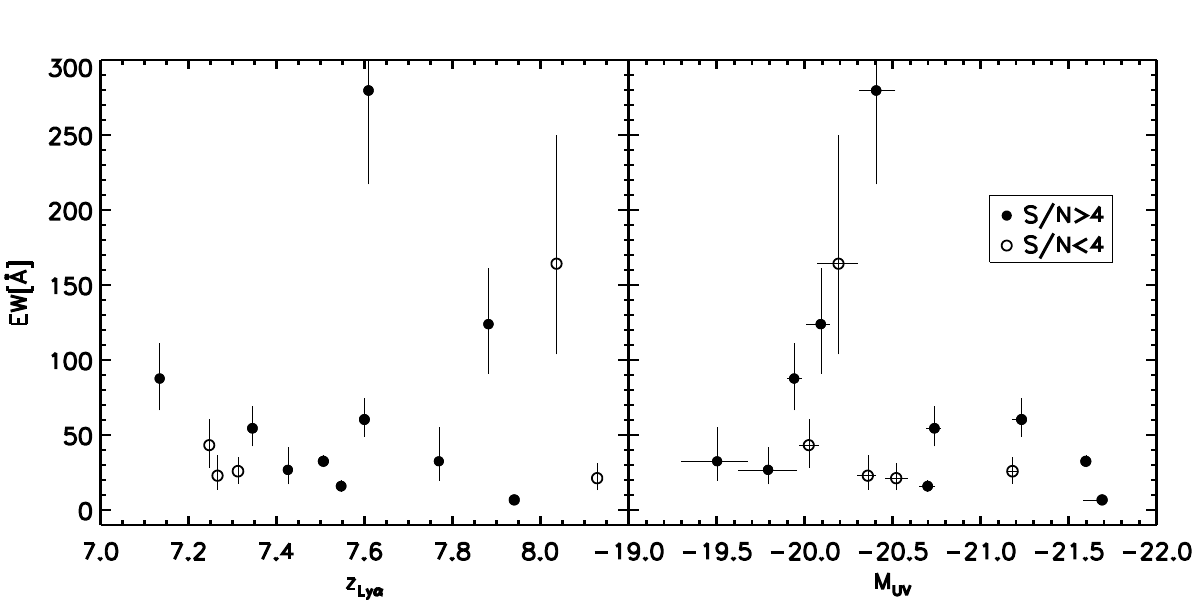}
\caption{Rest-frame EWs of Ly$\alpha$ emission lines vs. redshift (left) and $M_{\text{UV}}$ (right). Filled circles are S/N $>$ 4 LAEs, and the empty circles are 3 $<$ S/N $<$ 4 LAEs. We have 6 large-EW ($>$50\AA) LAEs, which includes a $z=8.04$ LAE detected with S/N $=3.9$. Their $M_{\text{UV}}$ values span down to $-21.2$; thus, it appears that such large-EW LAEs are not limited to UV-faint objects. Interestingly, we detect one LAE with the extremely large EW of 280\AA\ at $z=7.61$ without other AGN signatures (e.g., \ion{N}{5} emission), which might require a very young and massive metal-poor stars.}
\label{fig:lae_ew}
\end{figure*}

\subsubsection{Spectroscopic Confirmation of Galaxies at $z>7$}
Our highest redshift Ly$\alpha$ emission line with S/N$>$4 is detected at $z_{\text{spec}}=7.94$ (z7\_GND\_10402). Overall, we detected 10 significant Ly$\alpha$ emission lines above a 4$\sigma$ level at $z>7$, discovering 5 new Ly$\alpha$ emission lines at $z>7.5$ with 2 previously known Ly$\alpha$ emitters at $z_{\text{spec}}=7.51$ and $7.60$ \citep{Finkelstein2013a, Jung2019a}.  This increases the current number of confirmed Ly$\alpha$ emitters at $z>7.5$ from 10 to 15.  Specifically, the spectroscopic confirmation rate at a $4\sigma$ level is $\sim$16\% (10 Ly$\alpha$-confirmed over 62 $z_{\text{phot}} > 6$ targets), which is achieved with $\sim$8hr of deep integration time per target on average.  Although a spectroscopic confirmation rate strongly depends on target selection (photometric redshift PDFs and $M_{UV}$ distribution) in addition to the Ly$\alpha$ transmission in the IGM, this suggests that deep integrations make Ly$\alpha$ detections possible for many objects even out to $z \sim$ 8.

\subsubsection{Large-EW Ly$\alpha$ Emission Lines}
We derive the rest-frame EWs of the detected Ly$\alpha$ lines, listed in Table \ref{tab:MOSFIRE_LAE_spectra}. The UV-continuum flux density for calculating EWs is the averaged flux density over a 1230--1280\AA\ window of the best-fit SED model. Figure \ref{fig:lae_ew} shows the EWs vs. redshift (left) and $M_{\text{UV}}$ (right).  As presented in Table \ref{tab:MOSFIRE_LAE_spectra} and Figure \ref{fig:lae_ew}, we have six LAEs with EW $>$ 50\AA\, which includes a $z=8.04$ LAE detected with S/N $=$ 3.9 \cite[also including the $z=7.6$ LAE in][]{Jung2019a}, while previous measurements reported a deficit of these high-EW ($>$50\AA) LAEs at $z>7$ \citep[e.g.,][]{Tilvi2014a}.  Along with the recent studies that find large-EW LAEs at $z\sim7.5$ \citep{Larson2018a, Jung2019a}, our discovery of five additional large EW LAEs implies that such large-EW LAEs are less rare at this redshift than previously expected.  Furthermore, their $M_{\text{UV}}$ values span down to $-21.2$; thus, it appears that such large-EW($>$50\AA) LAEs are not limited to UV-faint objects, although no EW $\gtrsim100$\AA\ LAE is detected from UV-brighter ($M_{\text{UV}}<-20.5$) galaxies. To investigate further, we require a statistical number of LAEs.

Interestingly, z7\_GND\_34204 is emitting Ly$\alpha$ with EW = 280\AA\ comparable to the typical theoretical limit ($\sim$ 240 -- 350\AA) of Ly$\alpha$ emission from star formation \citep{Schaerer2003a}.  Although it has been suggested that Ly$\alpha$ fluorescence illuminated by a nearby quasar could contribute to large-EW Ly$\alpha$ \citep{Cantalupo2012a, Rosdahl2012a, Yajima2012a}, we did not detect an indicator of active galactic nucleus (AGN) activity from z7\_GND\_34204, for instance, a significant \ion{N}{5} emission. Also, z7\_GND\_34204 has $M_{\text{UV}}=-20.4$, comparable but a little fainter than the characteristic population of $z\sim7-8$ galaxies of $M_{\text{UV}}\sim-21$; thus, it seems that its large Ly$\alpha$ EW is less likely due to the AGN activity.  Although it is challenging to interpret such extremely large-EW Ly$\alpha$ without AGN, \cite{Kashikawa2012a} argue that their extremely large-EW Ly$\alpha$ emitter requires a very young and massive metal-poor stars like Population III stars \citep[see also ][]{Schaerer2003a, Raiter2010a}. Furthermore, \cite{Santos2020a} reported a significant number of non-AGN LAEs with extreme EWs ($>$ 240\AA) at $z\sim2-6$. With all that being said, there seems to be an increasing demand for extreme stellar populations to explain these extremely large-EW LAEs. 

\subsubsection{Asymmetric Line Profile of Ly$\alpha$ Emission}
Asymmetric line profiles are theoretically expected due to a combined effect of the ISM and IGM absorption, although complex Ly$\alpha$ radiative processes make it difficult to interpret the observed Ly$\alpha$ lines \citep[e.g.,][]{Dijkstra2014b}. Within an optically thick medium, Ly$\alpha$ photons suffer resonant scattering with the \ion{H}{1} gas, which redistributes the frequencies of the photons, shaping the line spectra into double-peaked profiles with an extremely opaque line center.  In an outflowing medium, the emerging line profile has a stronger red peak than the blue peak due to the ISM kinematics, showing an asymmetric profile with sharp edges near the Ly$\alpha$ line center and extended red tails at their far sides \citep[e.g.,][]{Verhamme2006a}. Specifically, the front side of the outflowing gas, which is moving toward us, has a relative velocity close to the resonance to the blueshifted Ly$\alpha$ photons (shorter wavelength photons) whereas the redshifted photons (longer wavelength photons) are less likely go through the resonance to the front side of the outflowing gas. Conversely, the red-side photons are likely scattered back toward us, being resonantly scattered by the back side of the outflowing gas, which is moving away from us \citep[e.g.,][]{Dijkstra2014a}. 

Recent studies of Green Pea galaxies, which are often referred to as local analogs of high-$z$ LAEs, have allowed detailed analyses including the internal kinematics to be constrained \citep[e.g.,][]{Yang2016a, Yang2017a, Yang2017b, Verhamme2018a, Orlitova2018a}. At higher redshifts, asymmetric Ly$\alpha$ line profiles have been reported at $4<z<7$ \citep[e.g.,][]{Rhoads2003a,Dawson2007a, Ouchi2010a, Hu2010a, Kashikawa2011a, Mallery2012a, U2015a}, though detailed physical modeling has not been possible.  

In the high-$z$ universe, photons blueward of the Ly$\alpha$ line center are most likely absorbed by residual \ion{H}{1} gas in the IGM. The resulting spectrum thus has only an asymmetric red peak observable.  Additionally, into the epoch of reionization, the IGM absorption due to the damping wing optical depth could shape an asymmetric line profile with a sharp blue edge and an extended red tail \citep{Weinberger2018a}. Along with this theoretical expectation, recent observational studies reveal this asymmetric shape of a Ly$\alpha$ emission line at $z>7$ from their deep spectroscopic observations \citep{Oesch2015a, Song2016b, Jung2019a, Tilvi2020a} while \cite{Pentericci2018b} report an asymmetric Ly$\alpha$ line profile from their stacked analysis of Ly$\alpha$ emission lines at $z\sim7$.  However, the occurrence of double-peaked Ly$\alpha$ emission has also been reported at these redshifts up to $z\gtrsim6$ \citep{Hu2016a, Matthee2018a, Songaila2018a, Bosman2020a}. This suggests that a highly ionized and/or significantly inflowing medium could allow an escape of the blue-side Ly$\alpha$ photons.

Thanks to our deep spectroscopic observations along with \cite{Song2016b} and \cite{Jung2019a}, the analysis of the Ly$\alpha$ line profile is feasible for multiple sources. In Table \ref{tab:MOSFIRE_LAE_spectra}, we present the measured asymmetry ($\sigma_{\text{red}}$/$\sigma_{\text{blue}}$) of the detected Ly$\alpha$ line profiles in the last column. Our measurements show asymmetric profiles of most Ly$\alpha$ emission lines with narrower blue-side profiles (Log$(\sigma_{\text{red}}$/$\sigma_{\text{blue}})>0$) in all S/N$>$4 Ly$\alpha$ emitters (except for z7\_GND\_39781), although only significant at the $\sim$1$\sigma$--2$\sigma$ level. 

The FWHM values from the red side of the line profiles are listed in the second-to-last column, spanning from $\sim$90 -- 410 km s$^{-1}$ with the median FWHM of $\sim$ $270$ km s$^{-1}$. Although the individual FWHM values largely vary, the median is comparable to those from the stacked spectra of LAEs at $z\sim6-7$ from the previous studies: $270\pm16$ and $265\pm37$ km s$^{-1}$ at $z=5.7$ and $6.6$ from \cite{Ouchi2010a}, and $290\pm25$ and $215\pm20$ km s$^{-1}$ at $z\sim6$ and $7$ from \cite{Pentericci2018b}. 

\subsection{Photometric Redshift Calibration}
The photometric selection technique of high-$z$ galaxies has brought an extensive set of candidate galaxies based on multiwavelength imaging data \citep[e.g.,][]{Stark2009a, Papovich2011a, Bouwens2015a, Finkelstein2015a}, and multiobject spectroscopic observations with photometrically selected galaxies have successfully confirmed their redshifts up to $z\gtrsim8$ \citep[e.g.,][]{Zitrin2015a}. Future observations with JWST/NIRSpec will be able to observe numerous fainter galaxies during the epoch of reionization with a much wider wavelength coverage of 0.9 -- 5.0$\mu$m. This enables us to spectroscopically confirm their redshifts via a suite of rest-frame UV and optical lines as well as Ly$\alpha$ emission \citep{Finkelstein2019a}. However, photometric selection such as photometric redshift measurements \citep[e.g., EAZY;][]{Brammer2008a} becomes more uncertain and needs to be calibrated further at $z\gtrsim6$ \citep{Pentericci2018b}. Particularly, the so far rare spectroscopic confirmation of galaxies at $z>7.5$ makes it challenging to test photometric redshift measurement due to the faint nature of distant galaxies and the increasing neutral fraction of the IGM. 

\begin{figure}[t]
\centering
\includegraphics[width=1.05\columnwidth]{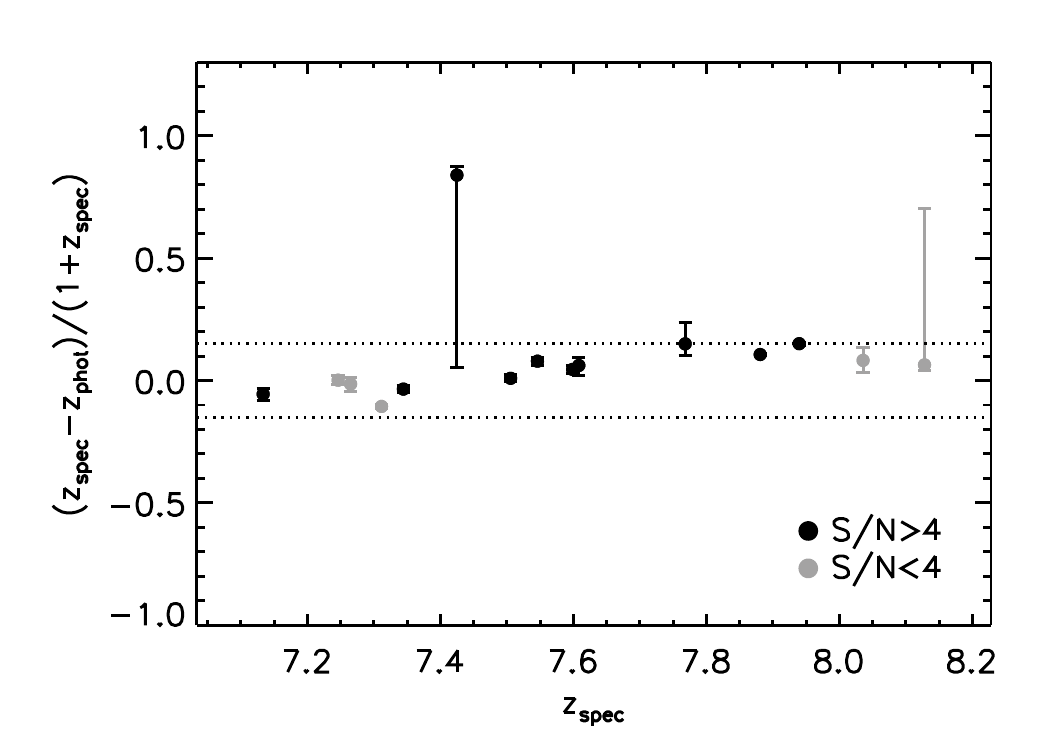}
\caption{A comparison between spectroscopic redshifts and photometric redshifts. The Y-axis is the relative $\Delta z = (z_{\text{spec}}-z_{\text{phot}})/(1-z_{\text{spec}})$, and the dotted lines indicate thresholds for the catastrophic outlier of $\Delta z>0.15$ as defined in \cite{Dahlen2013a} \citep[see also][]{Pentericci2018b}. The overall quality of the photometric redshifts appears good, and all the relative errors of $\Delta z$ are less than the defined outlier thresholds within their uncertainties. However, there is a systematic bias seen at $z>7.4$ where the photometric redshifts are always underestimated compared to their spectroscopic redshifts.}
\label{fig:specz_photz}
\end{figure}

Our comprehensive spectroscopic campaign now delivers new spectroscopic redshifts ($z_{\text{spec}}$) for 10 (15) galaxies at a 4$\sigma$ (3$\sigma$) level, including two previously known galaxies at $z=7.51$ and $7.60$ \citep{Finkelstein2013a, Jung2019a}. With these confirmed redshifts, we test the accuracy of our sample's photometric redshifts ($z_{\text{phot}}$) as shown in Figure \ref{fig:specz_photz}. We measure the relative error of  $\Delta z = (z_{\text{spec}}-z_{\text{phot}})/(1-z_{\text{spec}})$ and define outliers at $|\Delta z|>0.15$ similarly to \cite{Dahlen2013a}.  The overall quality of the photometric redshifts appears good, and all the relative errors of $\Delta z$ are less than the defined outlier thresholds within their uncertainties. However, there is a systematic bias at $z>7.4$ seen in the figure, where the photometric redshifts are always underestimated compared to the spectroscopic redshifts. A similar bias has been reported at various redshift ranges at $z>3$ in literature \citep[e.g.,][]{Oyarzun2016a, Brinchmann2017a, Pentericci2018b}. This is understandable in the sense that for Ly$\alpha$-detected objects, Ly$\alpha$ emission contributes to the increased flux at the photometric band which covers the observed wavelength range of Ly$\alpha$, and such increased flux pushes the continuum break (Lyman-break) toward shorter wavelengths in the photometric data. 

\begin{figure*}[t]
\centering
\includegraphics[width=0.68\paperwidth]{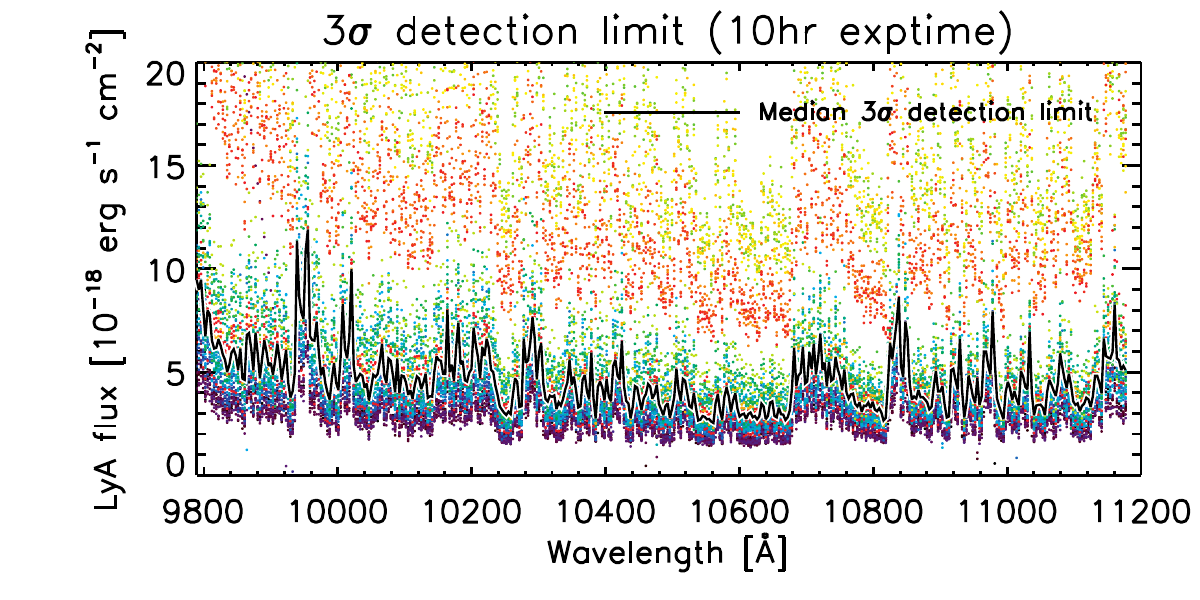}
\caption{The $3\sigma$ detection limit of an emission-line flux across the MOSFIRE instrument wavelength coverage, measured with 3\AA\ spacing using a Monte Carlo simulation, inserting mock emission lines.  We derive a linear relation between the line strength and its S/N level across the instrument wavelength coverage, and the detection limit of each simulated Ly$\alpha$ is interpolated from the precalculated linear relation. As the exposure time varies depending on each target, the detection limits are scaled by $\sqrt{t}$ to have 10hr of integration time for the purposes of this figure. The colored dots show the measured detection limit from the different galaxies, and the median detection limit is drawn as a black curve.  Between the sky emission lines, the typical $3\sigma$ detection limit is as low as $\sim$ 2--3 $\times 10^{-18}$ erg s$^{-1}$ cm $^{-2}$ for the targets observed in good observing conditions while 17 targets under poor observing conditions (2014 April, 2014 May, 2015 February) show three to four times higher detection limits.}
\label{fig:mosfire_detection_limit}
\end{figure*}

\section{L\lowercase{y}$\alpha$ Equivalent-width Distribution}
\subsection{Measuring the Ly$\alpha$ EW Distribution at $z\sim7.6$}
The Ly$\alpha$ EW distribution is often described by an exponential form, $P(\text{EW})\propto\text{exp}^{-\text{EW}/W_0}$, where $W_0$ is the $e$-folding scale \citep[e.g.,][]{Cowie2010a}. As \cite{Treu2012a, Treu2013a} suggested, a Ly$\alpha$ study as a probe of reionization benefits from using the Ly$\alpha$ EW distribution (over the more traditional Ly$\alpha$ fraction) as it includes more information such as the Ly$\alpha$ flux and UV-continuum brightness, in addition to the detection rate \citep[e.g.,][]{Tilvi2014a, Mason2018a, Jung2018a, Hoag2019a}. In \citet[][J18 hereafter]{Jung2018a}, we introduced a new methodology of measuring the Ly$\alpha$ EW distribution. We developed a simulation, which constructs a template of an expected number of Ly$\alpha$ emitters as a function of detection significance, by accounting for all types of data incompleteness, such as instrumental wavelength coverage, the wavelength-dependent Ly$\alpha$ detection limit, the UV-continuum flux, and the photometric redshift PDF. Here we apply this scheme to our MOSFIRE dataset to measure the Ly$\alpha$ EW distribution at $z>7$. As the MOSFIRE $Y$-band throughput allows us to detect Ly$\alpha$ at $7.0<z<8.2$, we limit the redshift range as $7.0<z<8.2$, placing our median constraint at $z\sim7.6$.

\begin{figure}[t]
\centering
\includegraphics[width=\columnwidth]{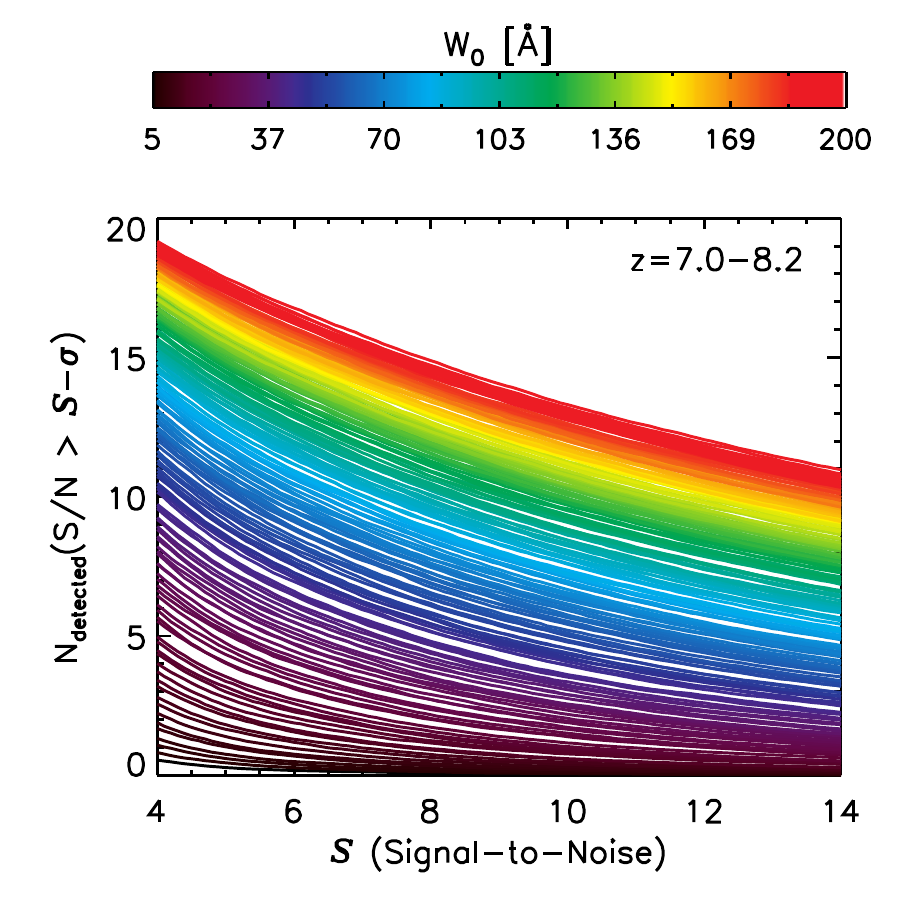}
\caption{The expected number of emission lines as a function of S/N level ($\mathcal{S}$) with various EW distributions ($W_0=$5--200\AA) at $z\sim7.6$. A larger choice of the $e$-folding scale ($W_0$) of the Ly$\alpha$ EW distribution (redder color) predicts a larger number of detected Ly$\alpha$ emission lines.}
\label{fig:MOS_n_detection}
\end{figure}

To predict the expected number of Ly$\alpha$ emitters for a given EW distribution, we first need to calculate the detection sensitivity for each object's spectrum.  We precompute these via MC simulations with the 1D spectra by adding a mock Ly$\alpha$ emission line to the reduced 1D spectra for each object and recover the line flux of this mock line with its error in the same manner as we performed for the detected Ly$\alpha$ emission lines. We create this mock emission line having an intrinsic line profile equal to the best-fit asymmetric Gaussian profile of our highest-S/N Ly$\alpha$ emission detected in z7\_GND\_42912. This is a reasonable choice as the emission line from z7\_GND\_42912 has a somewhat representative shape of the line profile with its FWHM ($\sim$9\AA) close to the median of all our S/N$>$4 emission lines. We estimate the Ly$\alpha$ detection sensitivity with 3\AA\ spacing for all observed targets individually (Figure \ref{fig:mosfire_detection_limit}). This detection sensitivity reflects observing conditions, instrument throughput, and sky emission lines. As the shape of the mock emission-line profile could affect the estimated emission-line sensitivity, we test a narrower choice of the mock emission-line profile with FWHM = 5\AA\ as well. Although its sharper profile provides a slightly lower detection limit, the overall difference of the line detection limit between FWHM = 5 and 9\AA\ does not exceed the $\sim$10\% level.

\begin{figure}[t]
\centering
\includegraphics[width=\columnwidth]{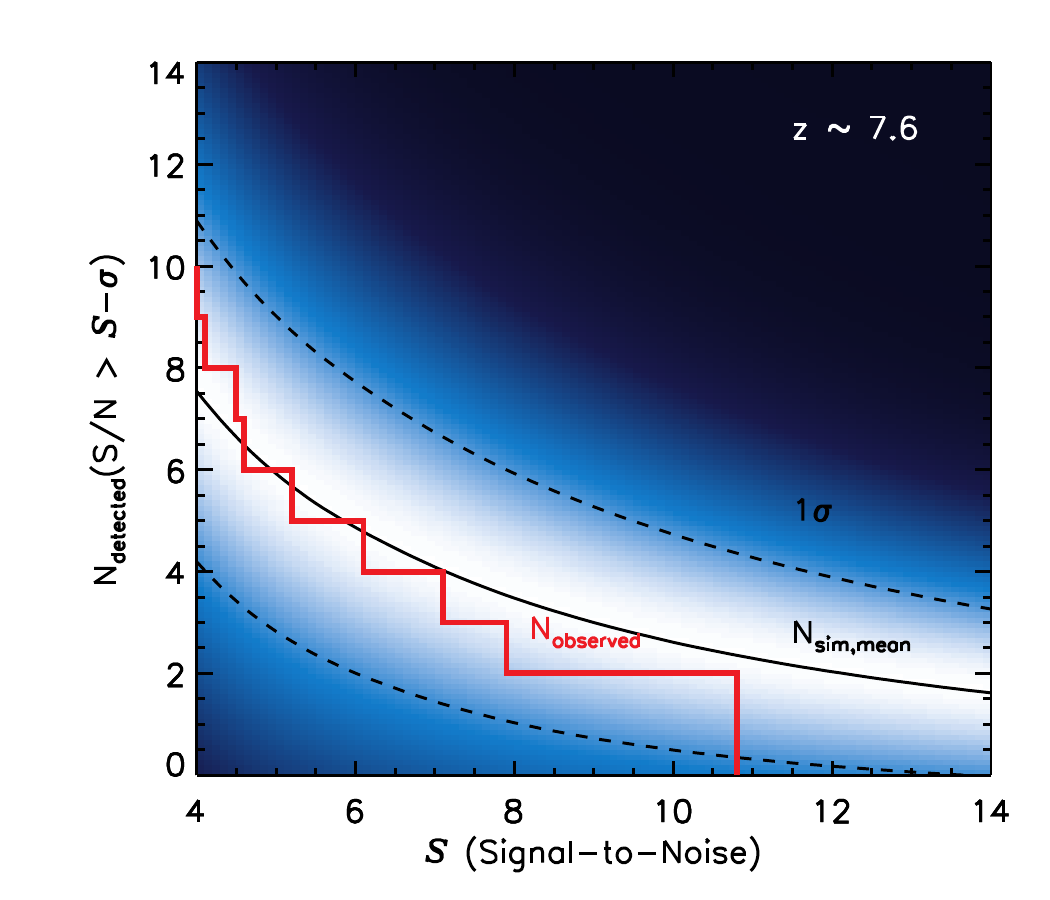}
\includegraphics[width=\columnwidth]{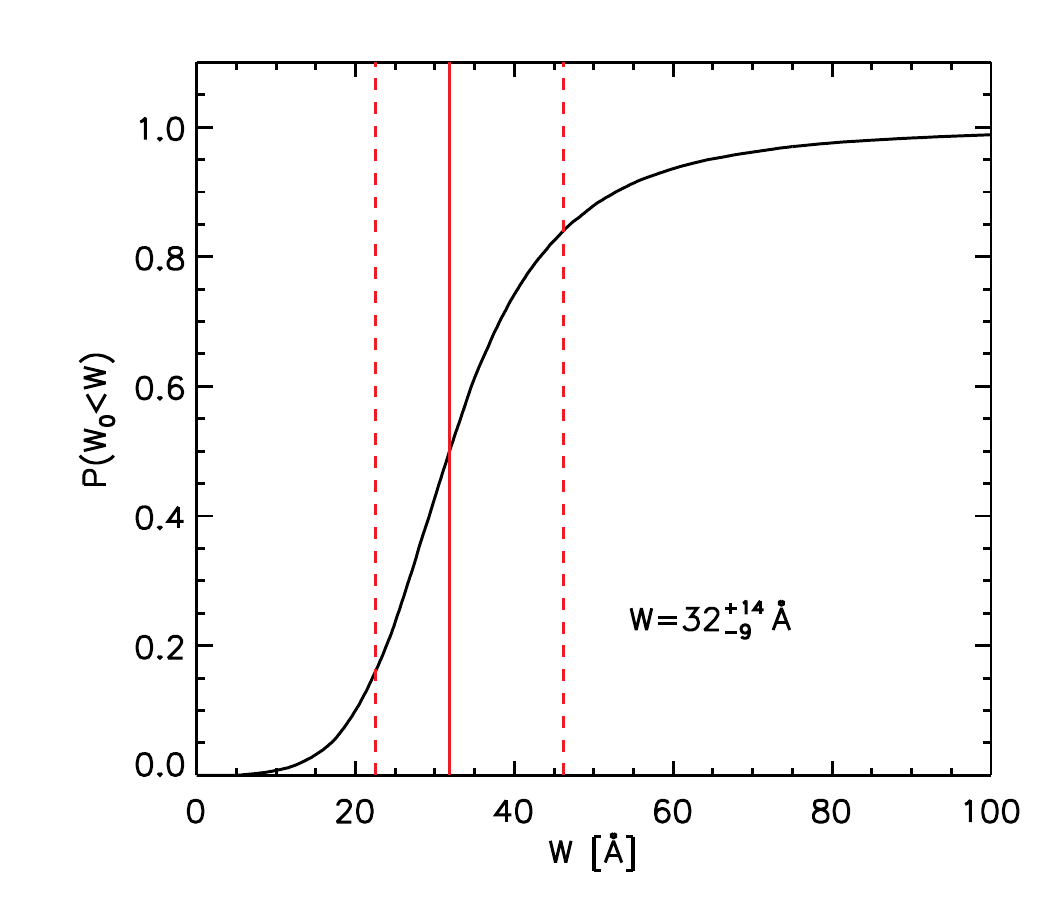}
\caption{(Top) The probability distribution of the expected number of Ly$\alpha$ emission lines as a function of S/N level ($\mathcal{S}$) at $z\sim7.6$, which is obtained from the $10^5$ MCMC chain steps.  Higher probability regions are denoted by the brighter colors.  The black solid curve shows the mean of the expected number of emission lines from our simulations as a function of S/N ($\mathcal{S}$), and the dashed curves are 1$\sigma$ uncertainties.  Our 10 emission lines are drawn a red solid line. (Bottom) The cumulative probability of the EW $e$-folding scale ($W_{0}$) from our MCMC-based fitting algorithm at $z\sim7.6$.  The median value and 1$\sigma$ boundaries are denoted with solid and dashed red vertical lines, respectively.}
\label{fig:MOS_w0}
\end{figure}

With these sensitivities in hand, there are then three main steps to simulate the expected number of Ly$\alpha$ emitters in our dataset: (i) allocate the wavelength of the simulated Ly$\alpha$ emission, which is randomly drawn from an object's photometric redshift PDF, (ii) estimate the line flux of the simulated Ly$\alpha$ by drawing an EW from the assumed Ly$\alpha$ EW distribution, $P(\text{EW})\propto\text{exp}^{-\text{EW}/W_0}$ given a value of $W_0$, multiplied by the UV-continuum flux of a galaxy, and (iii) calculate the S/N value of the simulation line using the precomputed wavelength-dependent Ly$\alpha$ detection limits.  We follow these steps to calculate the expected number of detected emission lines as a function of line S/N for a given value of the $e$-folding scale $W_0$, in the range of $W_0$ = 5 -- 200\AA. For each choice of $W_0$, we perform 1000 sets of simulations, which produce a distribution of the expected number of Ly$\alpha$ emission lines as a function of S/N, shown in Figure \ref{fig:MOS_n_detection}. In this figure, each curve is the median average of 1000 MC runs for a corresponding $W_0$. With a larger choice of $W_0$, more Ly$\alpha$ emission lines detected at higher-S/N levels would be expected in observations.

Lastly, we fit our actual Ly$\alpha$ emission lines (10 emission lines with S/N$>$4) to these simulated distributions to calculate the PDF of the $e$-folding scale ($W_0$) of the Ly$\alpha$ EW distribution. Our fitting scheme is based on a Markov Chain Monte Carlo (MCMC) algorithm with a Poisson likelihood, as counting the number of Ly$\alpha$ emission lines is a general Poisson problem. We use the "Cash statistic," which describes the Poisson likelihood \citep{Cash1979a}.  Via the Metropolis-Hastings MCMC sampling \citep{Metropolis1953a, Hastings1970a}, we construct the PDF of $W_0$, generating 10$^5$ MCMC chains.  A more detailed explanation of our fitting procedure is described in Section 4 in J18. 

\begin{figure*}[t]
\centering
\includegraphics[width=0.65\paperwidth]{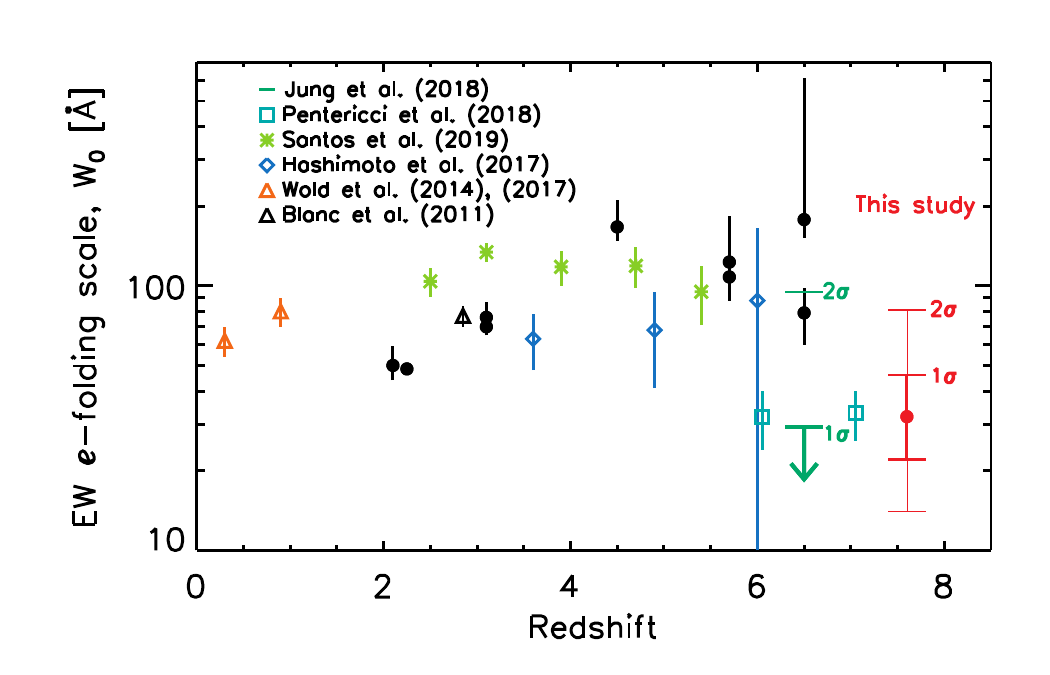}
\caption{The redshift dependence of the Ly$\alpha$ EW $e$-folding scale ($W_{0}$) up to $z\sim7.6$.  All data are shown without an IGM absorption correction.  Our $W_0$ measurements are denoted by a red filled circle at $7.0<z<8.2$.  Our study provides a decreased $W_0$ of $32^{+14}_{-9}$\AA\ at $z\sim7.6$, whereas there is little/no redshift evolution of $W_0$ reported in the literature at $z<6$. Black filled circles are low-$z$ measurements drawn from a compilation of \cite{Zheng2014a}, including \cite{Guaita2010a} at $z=2.1$, \cite{Nilsson2009a} at $z=2.25$, \cite{Gronwall2007a} at $z=3.1$, \cite{Ciardullo2012a} at $z=3.1$, \cite{Ouchi2008a} at $z=3.1, 3.7$, \cite{Zheng2014a} at $z=4.5$, \cite{Kashikawa2011a} at $z=5.7$ and $6.5$, and \cite{Hu2010a} at $z=5.7$ and $6.5$.  Blue diamonds are the measurements of \cite{Hashimoto2017a} at $z\sim3$--6 using the LAEs ($M_{\text{UV}}<-18.5$) from the MUSE HUDF Survey \citep{Bacon2017a}, which are consistent with \cite{Zheng2014a} at that redshift range.  At lower redshift, the $W_0$ measurements of \cite{Wold2017a} at $z\sim0.3$ and \cite{Wold2014a} at $z\sim0.9$ (orange triangles) suggest a relatively unevolving EW $e$-folding scale of Ly$\alpha$ across $z\sim0.3 - 3.0$, considering the other measurements described above, including \citet[black triangle]{Blanc2011a} at $z\sim2.85$. The green downward arrows at $z\sim6.5$ are the measurements of \cite{Jung2018a}. Recent measurements from \cite{Pentericci2018b} and \cite{Santos2020a} are shown as green squares at $z\sim6$ and $7$ and yellowish-green asterisks at $2<z<6$. The $z\sim6$ measurement from \cite{Pentericci2018b} is displayed at $z=6.1$ to avoid an overlap with the \cite{Hashimoto2017a} data point. The measurements from J18, this work, and \cite{Pentericci2018b} at $z\gtrsim6$ with the decreased values of $W_0$ are based on UV-continuum selection and show agreement within their uncertainties. However, there is some tension with other narrowband (NB) selection studies at the similar redshift \citep{Hu2010a, Kashikawa2011a}, and this is possibly due to sample selection biases between UV-continuum selection \citep[J18, this study, and][]{Pentericci2018b} and NB selection (the other studies in this figure).} 
\label{fig:MOS_ew_evolution}
\end{figure*}

Our MCMC sampling provides the PDF of $W_0$ at $7.0<z<8.2$, which is shown in the bottom panel of Figure \ref{fig:MOS_w0}. The median value of $W_0$ from the PDF is $32^{+14}_{-9}$\AA\ (68\% confidence level). The top panel of the figure presents a color-coded probability distribution of the expected number of Ly$\alpha$ emission lines at the corresponding S/N ($\mathcal{S}$) levels. The vertical axis shows the cumulative number of emission lines above $\mathcal{S}-\sigma$. The bright-shaded region shows the area of highest probability from our MCMC simulation, which matches the observed Ly$\alpha$ emission lines (red solid line).

\subsection{Redshift Dependency of the Ly$\alpha$ EW $e$-folding Scale}
J18 constrained the Ly$\alpha$ EW distribution, its characteristic $e$-folding scale ($W_0$), with a DEIMOS dataset at $z\sim6.5$. A comparison of that to lower-redshift measurements in the literature suggests a suppressed Ly$\alpha$ visibility with a measured 1$\sigma$ (2$\sigma$) upper limit of $e$-folding scale at $<$36\AA\ ($<$125\AA), providing a weak sign of an increasing \ion{H}{1} fraction in the IGM.

Here we add our data point at higher redshift, $7.0<z<8.2$, to our MOSFIRE observations, $W_0=32^{+14}_{-9}$\AA, as shown in Figure \ref{fig:MOS_ew_evolution}. The figure displays a compilation of measurements in the literature at all redshifts, including our new measurement of $W_0$ at $z>7$, shown as the red filled circle at $z\sim7.6$. Our result is lower than the $z <$ 6 observations ($W_0\sim$ 60--100\AA) at 1$\sigma$ confidence. This is expected, particularly into the epoch of reionization, with a more opaque IGM. However, we cannot rule out no evolution at the 2$\sigma$ level.

Compared to the J18 measurement showing a rapid drop at $z>6$ in its 1$\sigma$ upper limit (the green downward arrow in Figure \ref{fig:MOS_ew_evolution}), our $z\sim7.6$ measurement indicates a somewhat smoother decrease at $z>7$. This could be explained if reionization is inhomogeneous with regional variations in the IGM neutral fraction, though again these differences are not highly significant due to the somewhat large uncertainties.

J18 discuss some tension between their $z\sim6.5$ measurement and other narrowband (NB) selection studies at the same redshift \citep{Hu2010a, Kashikawa2011a}, mentioning possible sample selection biases between UV-continuum selection \citep[J18, this study, and][]{Pentericci2018b} and NB selection (the other studies in Figure \ref{fig:MOS_ew_evolution}). Interestingly, the constrained $e$-folding scale values at $z\sim6.5$ and $7.6$ from J18 and this study are mostly consistent with those at $z\sim6$ and $7$ from \cite{Pentericci2018b}. This agreement between continuum selection studies (and tension with the NB selection studies) might reflect possible biases caused by continuum selection where it misses large-EW LAEs from UV-faint galaxies. Such bias will be discussed in depth in the following section.

\subsection{The $W_0$ Dependence on $M_{\text{UV}}$}
The intrinsic shape of the Ly$\alpha$ EW distribution is known to be UV-magnitude dependent, and in general UV-bright galaxies have low EWs in Ly$\alpha$ \citep[e.g.,][]{Ando2006a, Stark2010a, Schaerer2011a, Cassata2015a, Furusawa2016a, Wold2017a, Hashimoto2017a, Jung2018a}. Additionally, \cite{Santos2020a} present no significant redshift evolution of the Ly$\alpha$ EW at $W_0=129^{+11}_{-11}$ using the full sample of SC4K \citep{Sobral2018a} LAEs at $z\sim2$ -- 6, but find a strong $W_0$ dependency on $M_{\text{UV}}$ and stellar mass. Thus, we need to be careful to interpret the redshift dependence of the EW distribution, as sample selection of spectroscopic observations would place a bias on the derived $W_0$. Our MOSFIRE observations mostly targeted galaxies $M_{\text{UV}}\lesssim-19.5$ at $z>7$, missing significant UV-faint populations. This could impact on our derived $W_0$, biasing it toward a smaller value \citep[e.g.,][]{Oyarzun2017a, Hashimoto2018a}.

With the $M_{\text{UV}}$ dependency in mind, it is critical to perform a fair comparison of $W_0$ at the same $M_{\text{UV}}$ between different redshifts. We measure $W_0$ from different magnitude ranges and compare them to the lower-redshift values from \cite{Santos2020a}, which are summarized in Table \ref{tab:EW_comparison}. At all $M_{\text{UV}}$ ranges, we find that $W_0$ is significantly lower at $z\sim7.6$, although in the brightest bin ($-22<M_{\text{UV}}<-21$), its $1\sigma$ upper limit overlaps with the lower-redshift value. 

Importantly, recent studies have been reported a sign of different evolution of the Ly$\alpha$ EW in bright and faint objects into the epoch of reionization \citep[e.g.,][]{Zheng2017a, Mason2018b} whereas a decreasing $W_0$ with increasing UV-continuum brightness is seen at lower redshift \citep[e.g.,][]{Ando2006a, Stark2010a, Schaerer2011a, Cassata2015a, Furusawa2016a, Wold2017a, Hashimoto2017a, Oyarzun2017a, Santos2020a}. Our measurements in Table \ref{tab:EW_comparison} also show an apparent upturn of $W_0$ at the brightest magnitude bin, which is consistent with the other studies at this redshift.  However, as the errors are large on these measurements, the result is also consistent with no upturn at the 1$\sigma$ level.

\begin{deluxetable}{ccc}[t]
\tablecaption{Ly$\alpha$ EW $e$-folding scale ($W_0$) at Different $M_{\text{UV}}$\label{tab:EW_comparison}
} 
\tablehead{
\colhead{$M_{\text{UV}}$} & \colhead{\cite{Santos2020a}}   & \colhead{This Study}  \\ 
	\colhead{}  & \colhead{$z\sim2-6$} & \colhead{$z\sim7.6$}
}
\startdata
{$-20<M_{\text{UV}}<-19$} & {$178^{+13}_{-13}$\AA} & {$61^{+64}_{-33}$\AA} \\  
{$-21<M_{\text{UV}}<-20$} & {$73^{+10}_{-10}$\AA} & {$28^{+18}_{-9}$\AA} \\  
{$-22<M_{\text{UV}}<-21$} & {$54^{+11}_{-11}$\AA} & {$48^{+63}_{-26}$\AA} \\
\hline
{Full sample} & {$129^{+11}_{-11}$\AA} & {$32^{+14}_{-9}$\AA} \\
\enddata
\end{deluxetable}

\subsection{Intrinsic Ly$\alpha$ Emitter Fraction}
In our $W_0$ measurement, we simulate mock Ly$\alpha$ emission lines assuming all star-forming galaxies at this redshift are emitting Ly$\alpha$ as the galaxies are metal poor and contain less dust, which promotes the escape of Ly$\alpha$. However, it is not well known what fraction of LBGs at $z>7$ would be actually emitting Ly$\alpha$ if it were not absorbed by the IGM. Although the Ly$\alpha$ emitter fraction (LAF) increases with increasing redshift, it is below 50\% at $z\lesssim6$ \citep[e.g.,][]{Stark2011a, Curtis-Lake2012a, Mallery2012a}. Thus, the assumption with an LAF of 100\% may be too optimistic, even if the extrapolated LAF continuously increases at $z>6$.

Thus, we perform our $W_0$ measurement as described in Section 4.1 again, but assuming an intrinsic LAF of 50\%. This gives a roughly doubled $W_0$ value of $67^{+54}_{-27}$\AA\ from the entire sample: $83^{+71}_{-48}, 55^{+59}_{-26},$ and $100^{+65}_{-57}$\AA\ at $-20$$<$$M_{\text{UV}}$$<$$-19$, $-21$$<$$M_{\text{UV}}$$<$$-20$, and $-22$$<$$M_{\text{UV}}$$<$$-21$, respectively. Although the measured $W_0$ values are still reduced at this redshift relative to the values at $z<7$, it is critical to understand the intrinsic LAF during the epoch of reionization in the future as it dramatically changes the inferred Ly$\alpha$ transmission in the IGM (see discussion in Section 5.1). 

\section{Constraints on Reionization}
\subsection{IGM \ion{H}{1} Fraction Inference}
\subsubsection{Ly$\alpha$ Transmission in the IGM}
A key quantity that we can draw from our measurement of the Ly$\alpha$ EW distribution is the Ly$\alpha$ transmission in the IGM, $T^{\text{Ly$\alpha$}}_{\text{IGM}}(=$EW$_{\text{obs}}$/EW$_{\text{int}})$, which compares the observed EW distribution (EW$_{\text{obs}}$) to the intrinsic EW distribution (EW$_{\text{int}}$). However, EW$_{\text{int}}$ is not directly observable during the epoch of reionization as Ly$\alpha$ has likely been affected by \ion{H}{1} in the IGM. Instead, to describe the intrinsic EWs at $z>6$, we utilize Ly$\alpha$ EWs obtained at $z<6$ \citep[e.g.,][]{Santos2020a} where we can assume that the universe is completely ionized. Although other measurements at $z\sim6$ are available from \cite{De-Barros2017a} and \cite{Pentericci2018a}, the reduced EWs from these studies suggest that reionization was not completed by $z\sim6$, thus their EWs may be affected by residual \ion{H}{1} in the IGM.

We assume no/little evolution of the interstellar medium (ISM) and circumgalactic medium (CGM) between $z<6$ and $z>6$ in this study, although a goal for future work is to figure out how the Ly$\alpha$ transmission in the ISM and CGM evolves over time in more detail (e.g., the evolution of the intrinsic Ly$\alpha$ escape fraction). Specifically, as shown in Figure \ref{fig:MOS_ew_evolution}, a compilation of Ly$\alpha$ EW measurements in the literature suggests no/little evolution of the $e$-folding scale ($W_0$) of the EW distribution at $\sim$100\AA\ in the ionized universe at $z<6$ although the different sample selection methods and/or $M_{\text{UV}}$ dependency of $W_0$ make this difficult to interpret. 
With that in mind, this assumption allows us to separate the IGM attenuation from the ISM and CGM effect on the Ly$\alpha$. 

It is critical to take the $M_{\text{UV}}$ dependency into account when estimating $T^{\text{Ly$\alpha$}}_{\text{IGM}}$ as the measured $W_0$ shows a clear $M_{\text{UV}}$ dependence.  We thus utilize the $M_{\text{UV}}$-constrained $W_0$ values in Table \ref{tab:EW_comparison} and estimate $T^{\text{Ly$\alpha$}}_{\text{IGM}}(=$EW$_{z\sim\text{7.6}}$/EW$_{z\sim\text{2--6}})= 0.34^{+0.42}_{-0.19}, 0.38^{+0.35}_{-0.15}$, and $0.89^{+0.11}_{-0.55}$ at $-20<M_{\text{UV}}<-19$, $-21<M_{\text{UV}}<-20$, and $-22<M_{\text{UV}}<-21$, respectively. For the remainder of our analysis, we set $T^{\text{Ly$\alpha$}}_{\text{IGM}}$ = $0.38^{+0.35}_{-0.15}$ as our fiducial value, measured at $-21<M_{\text{UV}}<-20$ where the bulk of our spectroscopic sample lies (Figure \ref{fig:MOSFIRE_muv}).

\subsubsection{Ly$\alpha$ Optical Depth in the IGM}
From a theoretical perspective \citep[e.g.,][]{Dijkstra2014b, Mesinger2015a, Mason2018a, Weinberger2018a, Weinberger2019a}, $T^{\text{Ly$\alpha$}}_{\text{IGM}}$ is often described with an $e^{-\tau}$ modeling of the Ly$\alpha$ radiative transfer as 
\begin{equation}
T^{\text{Ly$\alpha$}}_{\text{IGM}}=\frac{\int d\nu J(\nu) e^{-\tau_{\text{IGM}}(\nu)}}{\int d\nu J(\nu)},
\end{equation}
where $J(\nu)$ is the intrinsic Ly$\alpha$ emission line from galaxies, and $\tau_{\text{IGM}}(\nu)$ is the IGM optical depth. The IGM optical depth, $\tau_{\text{IGM}}(\nu)$, is commonly considered as a combination of the damping wing optical depth due to diffuse \ion{H}{1} during reionization ($\tau_{D}$) and the optical depth due to resonant scattering within the CGM of galaxies ($\tau_{\text{HII}}$) as $\tau_{\text{IGM}}(\nu)=\tau_{\text{D}}(\nu)+\tau_{\text{HII}}(\nu)$ \citep[e.g.,][]{Dijkstra2014b}.  

Although one can model the CGM contribution ($\tau_{\text{HII}}$) realistically with a combined description of reionization models \citep[e.g.,][]{Weinberger2018a}, we take a simple approach by assuming no/little redshift variation of $\tau_{\text{HII}}$ at fixed $M_{\text{UV}}$ (or fixed halo mass) between $z<6$ and $z>6$.

\subsubsection{\ion{H}{1} Fraction in the IGM at $z\sim7.6$}
As an increasing neutral fraction ($X_{\text{HI}}$) in the IGM during the epoch of reionization determines the damping wing optical depth ($\tau_{D}$), the Ly$\alpha$ transmission in the IGM ($T^{\text{Ly$\alpha$}}_{\text{IGM}}$) is tied to $X_{\text{HI}}$, or the reionization history. A simplified analytical approach in \citet[][their Equation 30]{Dijkstra2014b} relates $X_{\text{HI}}$ with the damping wing optical depth as 
\begin{equation}
\tau_D(z_g,\Delta v)\approx 2.3X_{\text{HI}}\Big(\frac{\Delta v_{b}}{600 \text{km s}^{-1}}\Big)^{-1}\Big(\frac{1+z_g}{10}\Big)^{3/2},
\end{equation}
where $X_{\text{HI}}$ is the averaged neutral fraction of the IGM at a galaxy redshift $z_g$, and $\Delta v$ is the velocity offset of Ly$\alpha$ from the systemic redshift. $\Delta v_{b}$ represents a velocity offset from line resonance when a photon first enters a neutral cloud, written as $\Delta v_{b} = \Delta v + H(z_g)R_{b}/(1+z_g)$, where $H(z_g)$ is the Hubble expansion rate, and $R_{b}$ is the comoving distance to the edge of the neutral cloud. Initially, $\tau_{D}$ is primarily dependent on frequency, characterized by the line profile and the velocity offset \citep[also refer to Figure 5 in][for example]{Mason2018a}. Thus, the frequency dependency is transformed to a $\Delta v$ dependency in the equation. 

Once we relate the damping wing optical depth ($\tau_{D}$) to the estimated Ly$\alpha$ transmission of the IGM ($T^{\text{Ly$\alpha$}}_{\text{IGM}}$) following Equation (2), the primary uncertainties of the derived $X_{\text{HI}}$ in Equation (3) come from the velocity offset ($\Delta v$) and the distance of the first encounter to the closest neutral patch ($R_b$), which corresponds to the typical size of ionized bubbles at each redshift. Specifically, increasing $\Delta v$ and $R_b$ favors higher $X_{\text{HI}}$, which fosters an easier escape of Ly$\alpha$ photons before encountering a neutral IGM.

A typical observed range of the velocity offset at $z\gtrsim6$ is $\sim$100 -- 200 km s$^{-1}$ \citep{Stark2015a, Inoue2016a, Pentericci2016a, Mainali2017a, Hutchison2019a} while UV-luminous ($M_{\text{UV}}\lesssim-22$) systems reveal larger offsets \citep{Willott2015a, Stark2017a}. However, it is not feasible to directly measure $\Delta v$ from our observations without other metal lines detected. Instead, under the approximation that the FWHM of the Ly$\alpha$ emission equals the line velocity offset \citep{Yang2017a, Verhamme2018a}, we infer the velocity offsets from the FWHM values of our S/N $>4$ detected Ly$\alpha$ emission lines of $\sim$ 90 -- 410 km s$^{-1}$, using the mean value of $\sim$ 240 km s$^{-1}$ (Table \ref{tab:MOSFIRE_LAE_spectra}).

\begin{figure}[ht]
\centering
\includegraphics[width=1.0\columnwidth]{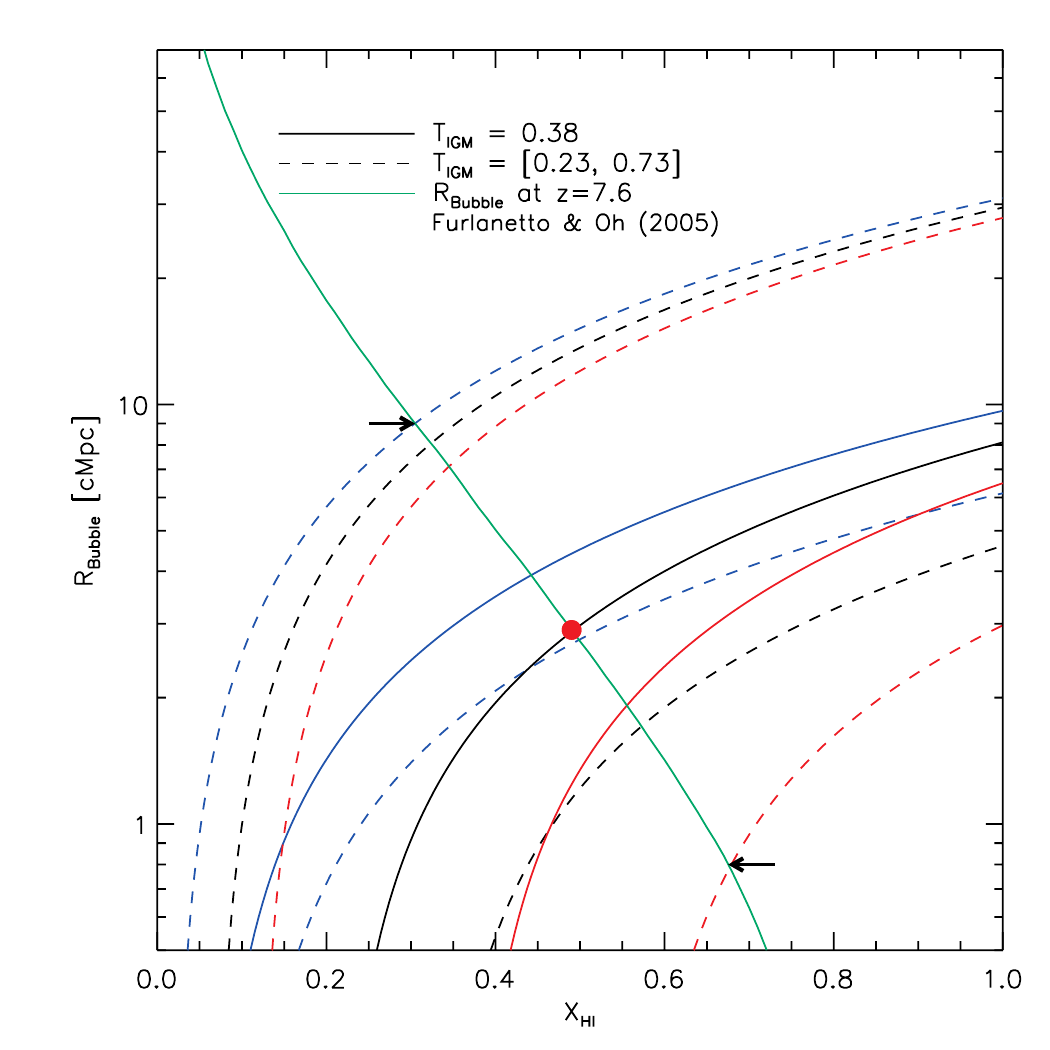}
\caption{IGM neutral hydrogen fraction ($X_{\text{HI}}$) measurement. The plot displays the characteristic size of ionized bubbles ($R_{b}$) at $z=7.6$ as a function of $X_{\text{HI}}$ (green line) from \cite{Furlanetto2005a} as well as our $X_{\text{HI}}$ estimates as a function of $R_{b}$ (other lines) from the analytic approach of \cite{Dijkstra2014b}. The bubble size ($R_{b}$) at $z=7.6$ is interpolated from the $z=6$ and $9$ values in Figure 1 of \cite{Furlanetto2005a}. In our $X_{\text{HI}}$ calculation, each line represents a different combination of the Ly$\alpha$ transmission in the IGM ($T^{\text{Ly$\alpha$}}_{\text{IGM}}$) and the velocity offset ($\Delta v$). We explore $1\sigma$ upper and lower limits of $T^{\text{Ly$\alpha$}}_{\text{IGM}}$ = [0.23, 0.73] (dashed lines) as well as the median value of $T^{\text{Ly$\alpha$}}_{\text{IGM}}=0.38$ (solid lines), allowing a range of $\Delta v$ from 70 km s$^{-1}$ (blue) to 420 km s$^{-1}$ (red) with the fiducial value of 240 km s$^{-1}$ (black). The red dot indicates the estimated $X_{\text{HI}}\sim49\%$ with the fiducial values of $T^{\text{Ly$\alpha$}}_{\text{IGM}}=0.38$ and $\Delta v=240$ km s$^{-1}$, simultaneously satisfying the predicted size of ionized bubble at a given $X_{\text{HI}}$ (corresponding to $R_b\sim2.9$ cMpc) from the reionization model. The black arrows indicate $1\sigma$ limits of $X_{\text{HI}}$ at $\sim30$ -- $68\%$, conservatively allowing a range of $\Delta v=70$--$420$ km s$^{-1}$. } 
\label{fig:X_HI}
\end{figure}

Although the damping wing optical depth ($\tau_{D}$) is strongly frequency dependent, dealing with the detailed modeling of the Ly$\alpha$ line profile is beyond the scope of this study. Instead, we assume no redshift evolution of the line profile in our analysis and adopt a representative value of the velocity offset from our observations. This is a reasonable approach under the assumption that  ISM conditions are similar at the same $M_{\text{UV}}$, also shown in the observed empirical relation between $\Delta v-M_{\text{UV}}$ \citep[e.g.,][]{Mason2018a}.

As mentioned, the $X_{\text{HI}}$ calculation in Equation (3) is dependent on the characteristic size of ionized bubbles ($R_b$) as well, which can be predicted by reionization models \citep[e.g.,][]{Furlanetto2005a, Mesinger2007a}. In Figure \ref{fig:X_HI}, the green line shows the characteristic bubble size at $z=7.6$ as a function of $X_{\text{HI}}$ based on the analytic model of reionization from \cite{Furlanetto2005a}. The bubble size ($R_{b}$) at $z=7.6$ is interpolated from the $z=6$ and $9$ values in Figure 1 of \cite{Furlanetto2005a}. The expected size of the bubbles in a highly ionized universe (low $X_{\text{HI}}$) would be larger than that in a more opaque universe (high $X_{\text{HI}}$). In Figure \ref{fig:X_HI}, we also plot our $X_{\text{HI}}$ estimates based on Equation (3) as a function of $R_{b}$. Each line represents a different combination of the Ly$\alpha$ transmission in the IGM ($T^{\text{Ly$\alpha$}}_{\text{IGM}}$) and the velocity offset ($\Delta v$): the median $T^{\text{Ly$\alpha$}}_{\text{IGM}}$ (solid) and its 1$\sigma$ limits (dashed), and $\Delta v=$ 70, 240, and 420 km s$^{-1}$ (blue, black, and red). The red dot indicates our fiducial value of $X_{\text{HI}}\sim49\%$ with $T^{\text{Ly$\alpha$}}_{\text{IGM}}=0.38$ and $\Delta v=240$ km s$^{-1}$ (from the mean FWHM of the Ly$\alpha$ emission lines). This simultaneously satisfies the predicted size of the ionized bubble at a given $X_{\text{HI}}$ (corresponding to $R_b\sim2.9$ cMpc) from the reionization model. In the figure, the black arrows indicate the $1\sigma$ limits of our $X_{\text{HI}}$ calculation at $\sim30 - 68\%$, conservatively allowing a range of $\Delta v=70$--$420$ km s$^{-1}$. 

The preferred size of the ionized bubbles, which satisfies the estimated $X_{\text{HI}}$, ranges from $\sim1$--9 cMpc. Compared to a numerical model \citep{Mesinger2007a}, this analytic model of \cite{Furlanetto2005a} slightly underestimates the size of ionized bubbles as it does not take overlapping bubbles into account. However, the preferred bubble size here ($\sim1$--9 cMpc) is still comparable to that predicted in \cite{Yajima2018a} where they calculate the size of \ion{H}{2} regions created by LAEs at $z\sim8$ through semianalytic modeling: $\sim2$--9 cMpc from $M_{\text{star}} = 10^{8 - 10} M_{\odot}$ galaxies (see their Figure 10).  Additionally, it is worth discussing the chance of detecting double-peaked Ly$\alpha$ emission from our LAEs as the recent discoveries of double-peaked Ly$\alpha$ emission at $z\gtrsim6$ \citep{Hu2016a, Matthee2018a, Songaila2018a, Bosman2020a} suggest that LAEs that reside in a highly ionized region could present double-peaked Ly$\alpha$ emission. Specifically, the preferred ionized bubble size from our calculation seems larger than the estimated bubble size of COLA1, a double-peaked LAE at $z=6.593$, which could form its $\sim$2.3 cMpc ionized bubble \citep{Matthee2018a}.  However, none of our LAEs at $z>7$ presents a significant sign of a double-peaked profile.  This implies that the escape of blue-side photons of Ly$\alpha$ appears less feasible at $z>7$ possibly with an increasing IGM neutral fraction.

Although our $X_{\text{HI}}$ inference is inevitably sensitive to a reference value of the Ly$\alpha$ EW distribution at $z<6$ and reionization models, our inferred $X_{\text{HI}}\sim49^{+19}_{-19}\%$ is lower in modest tension ($>$1$\sigma$) with results at similar redshifts: $X_{\text{HI}}=0.88$ at $z=7.6$, $> 0.76$ at $z\sim8$ \citep{Hoag2019a, Mason2019a} and more comparable to the lower-redshift measurement of $X_{\text{HI}}=0.55$ at $z\sim7$ \citep{Mason2018a, Whitler2020a}. This could be due to the intrinsic inhomogenous nature of reionization.  However, our result is consistent with what is predicted in \cite{Finkelstein2019b} of $X_{\text{HI}}$ $\sim$ $0.3^{+0.1}_{-0.1}$ at $z=7.6$, under the assumption that the faintest galaxies dominate the ionizing photon budget.  It is also consistent with the neutral fraction of $X_{\text{HI}}=39^{+22}_{-13}\%$, which is estimated from the damping wing analysis of a luminous $z=7.5$ quasar observation in \cite{Yang2020a}.  Additionally, as discussed in Section 4.4, the estimated $T^{\text{Ly$\alpha$}}_{\text{IGM}}$ would be doubled with a lower intrinsic LAF of 50\%, making the corresponding $X_{\text{HI}}$ values even lower. 

\begin{figure*}[hbt!]
\centering
\includegraphics[width=1.0\columnwidth]{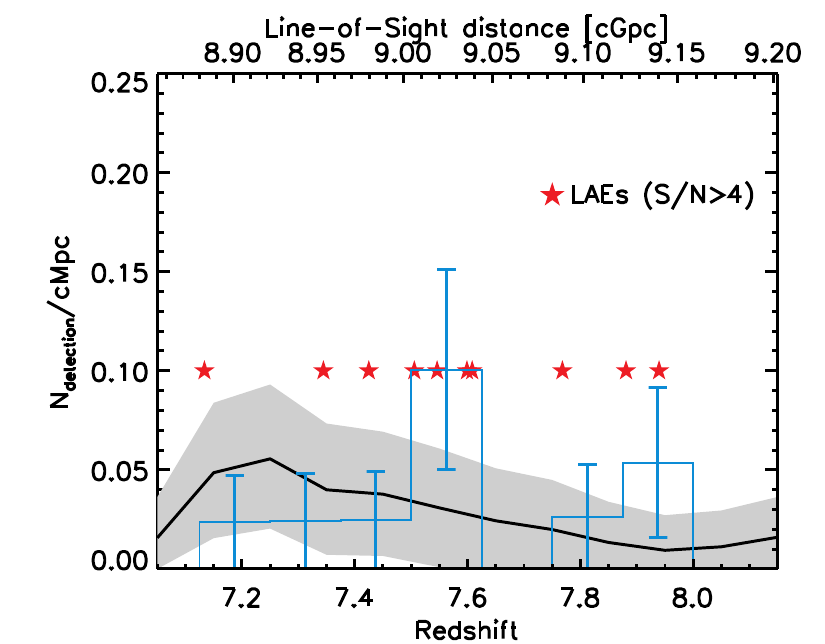}
\includegraphics[width=1.0\columnwidth]{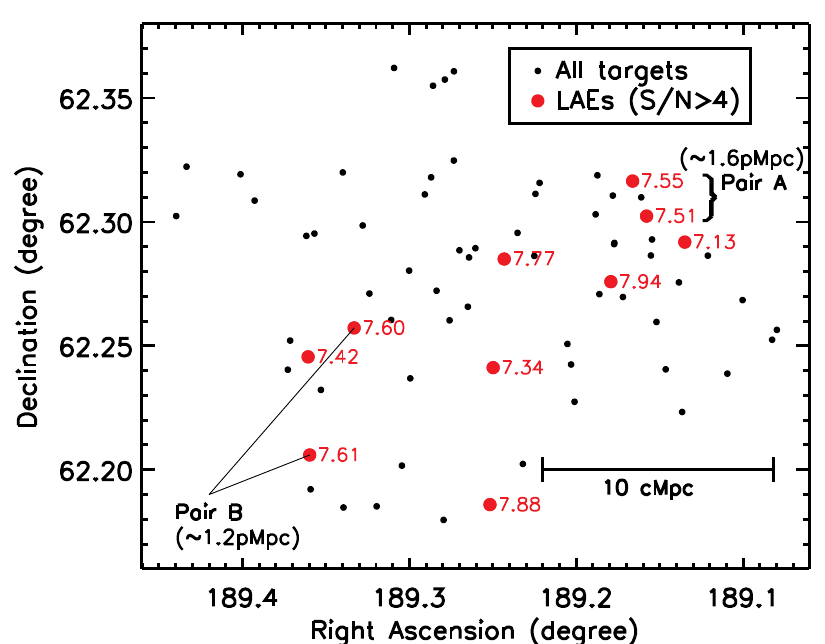}
\caption{(Left) LAE overdensity at a function of redshift (or line-of-sight distance). The y-axis is the number of Ly$\alpha$ emission lines ($N_{\text{detection}}$) per unit volume, a 1 cMpc-thick slice in the line-of-sight (LOS) direction. The black curve is the expected number of emission lines based on our EW distribution modeling. The blue histogram shows actual Ly$\alpha$ emission lines, and its error bar is obtained with Poissonian statistics as $\sqrt{N_{\text{detection}}}$ in each bin. The red stars denote the individual spectroscopic redshifts of the detected LAEs. A notable feature is the peak near $z=7.5 - 7.6$ where we detect more LAEs than expected. Four LAEs at $z=7.51, 7.55, 7.60$, and $7.61$ are clustered within $\Delta z\sim0.1$ (or 32.9 cMpc LOS distance). (Right) The spatial distribution of the detected LAEs (red circles) and all targets (black dots). The numbers next to red circles are the spectroscopic redshifts of the LAEs. The four clustered LAEs near $z=7.5 - 7.6$ are spread across the observed area, but still within 22.1 cMpc in the transverse direction. Particularly, two LAEs at $z=7.51$ and $7.55$ and the other two at $z=7.60$ and $7.61$ form close pairs, noted as ``Pair A" and ``Pair B" with their physical separations in parentheses in the plot. These clustered LAEs may indicate a sign of a large (with a $\sim$ 40 cMpc spatial extent) highly ionized structure (or multiple smaller ionized bubbles with the LAE pairs) in the early universe, showing directly the inhomogeneity of reionization.} 
\label{fig:lae_clustering}
\end{figure*}

\subsection{Ionization Structure of the IGM}
Reionization is an inhomogeneous process, starting in small ionized bubbles of the IGM around ionizing sources, the first stars and galaxies, with these bubbles expanding outward until the hydrogen in the IGM was completely ionized. The predicted size of these ionized bubbles thus becomes smaller with increasing redshifts \citep[see the review of][]{McQuinn2016a}. 

Recent observations show growing evidence of ionized bubbles at $z>7$. \cite{Zheng2017a} studied Ly$\alpha$ luminosity function (LF) from the Lyman Alpha Galaxies in the Epoch of Reionization Survey and reported a bump at the bright end of the Ly$\alpha$ LF at $z\sim7$. They suggest that this is indicative of large ionized bubbles ($>$1 cMpc radius) and also different evolution between the bright and faint ends of the Ly$\alpha$ LF. Additionally, \cite{Castellano2018a} presented a triplet of spectroscopically confirmed LAEs at the same redshift ($z=7.008$), which includes a pair of them at only $\sim$90 kpc distance. More recently, \cite{Tilvi2020a} reported spectroscopic confirmation of three galaxies likely in a group (EGS77) at $z=7.7$ within $<0.7$ Mpc physical separation, forming up to $\sim$1 pMpc-size ionized bubbles. 

Given our largest number of spectroscopically confirmed Ly$\alpha$ emitters from our survey, we explore spatial clustering of our detected LAEs at these high redshifts, and therefore, the inhomogeneity in the IGM. The left panel of Figure \ref{fig:lae_clustering} presents a comparison between the number of detected LAEs (blue histogram) and the number of expected LAEs ($N_{\text{exp}}$: black solid line) from our survey as a function of redshift. The y-axis is $N_{\text{exp}}$ per 1 cMpc-thick slice in the line-of-sight (LOS) direction over the entire survey area, which is calculated as described in Section 4.1, assuming $W_0=32$\AA. The shaded region represents the 1$\sigma$ uncertainty on $N_{\text{exp}}$, and the red stars denote the spectroscopic redshifts of the detected LAEs.  A notable feature is the peak near $z=7.5$--$7.6$ where we detect more LAEs than expected, whereas we have Ly$\alpha$-detected objects which are less than/comparable to $N_{\text{exp}}$ in other redshift bins.  Four LAEs at $z=7.51, 7.55, 7.60$, and $7.61$ are clustered within $\Delta z\sim0.1$ (or 32.9 cMpc LOS distance). The right panel displays the 2D spatial distribution of our target galaxies (black dots) and the LAEs (red dots). The four clustered LAEs near $z=7.5 - 7.6$ are spread across the observed area, but still within 22.1 cMpc in the transverse direction (projection on the sky), and spread over a 3D spatial extent of 39.6 cMpc.

Particularly, z7\_GND\_42912 at $z=7.51$ and z7\_GND\_6330 at $z=7.55$ are in close proximity with each other with a 1.55 pMpc physical separation, marked as ``Pair A" in the right panel of Figure \ref{fig:lae_clustering}.\footnote{To clarify, our discussion on LAE pairs in this section must be distinguished from the conventional definition of galaxy pairs in the context of galaxy--galaxy interactions. Instead, we discuss LAE pairs that overlap their individual ionized bubbles each other, forming contiguous ionized areas.}   Along the LOS, the two galaxies are separated by 1.53 pMpc while in the transverse direction, they are separated by a mere 52\farcs7 (0.27 pMpc). Also, the other two galaxies (z7\_GND\_16863 and z7\_GND\_34204; ``Pair B") form a close pair at $z=7.60$ with a 1.15 pMpc physical separation. They are separated by only 0.35 pMpc along the LOS and by 3\farcm2 (0.95 pMpc) in the transverse direction.

Spectroscopic confirmation of $z>7$ galaxies via Ly$\alpha$ emission implies that these galaxies must be surrounded by ionized bubbles, making them visible in Ly$\alpha$ emission. Referring to the models from \cite{Yajima2018a}, and following \cite{Tilvi2020a}, the sizes of individual \ion{H}{2} bubbles created by the galaxies could be roughly up to $\sim$1 pMpc. Specifically, z7\_GND\_42912 at $z=7.51$ (in Pair A) and z7\_GND\_16863 at $z=7.60$ (in Pair B) are massive ($M_{*}>10^9M_{\odot}$) and bright in their UV ($M_{\text{UV}}=-21.6$ and $-21.2$), which could be enough to form $\sim$1 pMpc-size ionized bubbles around them.  As listed in Table \ref{tab:MOSFIRE_LAE_spectra}, their Ly$\alpha$ luminosities are also bright ($>$10$^{43}$ erg s$^{-1}$) enough to form $\sim$1 pMpc-size ionized bubbles, based on the \cite{Yajima2018a} model (see their Figure 15).  Furthermore, z7\_GND\_34204 could form its largest $\sim$1.4 pMpc ionized bubble, suggested by its large EW ($=279.7$\AA) and bright Ly$\alpha$ luminosity ($=3.26\times10^{43}$ erg s$^{-1}$). Although our bubble size estimation is still model dependent, given their small separations in the two pairs ($\sim$1.6 pMpc in Pair A and $\sim$1.2 pMpc in Pair B), the individual \ion{H}{2} bubbles in each pair likely overlap, forming a contiguous ionized region.

Relating to the \ion{H}{1} fraction ($X_{\text{HI}}$) presented in the previous section, our lower $X_{\text{HI}}\sim49\%$ is certainly driven by this potentially large ionized structure, and higher $X_{\text{HI}}$ values from other studies could be similarly driven by neutral regions. For instance, if we exclude the four clustered LAEs, the inferred $X_{\text{HI}}$ is increased to $58^{+23}_{-16}\%$ with $T^{\text{Ly$\alpha$}}_{\text{IGM}}$ = $0.22^{+0.21}_{-0.10}$ at $-21<M_{\text{UV}}<-20$, which is still lower than, yet more comparable to, other studies. 

These clustered LAEs may indicate a sign of a  large (with a $\sim$ 40 cMpc spatial extent) highly ionized structure (or multiple smaller ionized bubbles with the LAE pairs) in the early universe. Thus, our observations of the ionization structure provide an increasing body of evidence of inhomogeneous reionization caused by individual or a group of galaxies during the middle phase of the reionization epoch. Such inhomogeneities of reionization presented here demonstrate that future studies with much wider-field Ly$\alpha$ surveys will better constrain the global evolution of reionization.

\section{Summary}
We carried out our analysis on a comprehensive Ly$\alpha$ spectroscopic survey dataset with Keck/MOSFIRE at $z>7$, a subset of the Texas Spectroscopic Search for Ly$\alpha$ Emission at the End of Reionization survey \citep{Jung2018a, Jung2019a}. We reduced 10 nights of Keck/MOSFIRE observations, targeting 72 high-$z$ galaxies in the GOODS-N field with deep exposure times of $t_{\text{exp}}=4.5$ -- 19 hr. Utilizing an improved automated emission-line search, we detect 10 Ly$\alpha$ emission lines at $z>7$ with $>$ 4$\sigma$ significance, and 5 more detections at a 3 -- 4$\sigma$ level. By simulating the expected number of Ly$\alpha$ emission lines in our targets, we constrain the Ly$\alpha$ EW distribution at $7.0<z<8.2$ with the detected Ly$\alpha$ lines and infer the IGM \ion{H}{1} fraction based on the Ly$\alpha$ transmission in the IGM. Also we study the spatial clustering of the LAEs to search for ionized structures during the epoch of reionization. Our major findings are summarized as follows.

\begin{enumerate}
\item We perform an automated search scheme on both 1D and 2D spectra to search in an unbiased way for plausible emission-line features, utilizing the automated 1D search algorithm of \cite{Larson2018a} and the Source Extractor software \citep[SExtractor;][]{Bertin1996a} on 2D spectra. Our automated search guarantees machine-driven consistency on detecting emission lines on 1D and 2D spectra, which supplement our visual inspection by discovering three more emission lines at the 4$\sigma$ level.

\item We detect 10 $z>7$ Ly$\alpha$ emission lines with S/N $>$ 4 and 7 at $z>7.5$, which includes our highest redshift Ly$\alpha$ emission with S/N $>$ 4 at $z_{\text{spec}}=7.94$. This significantly increases the total number of confirmed Ly$\alpha$ emission lines in this epoch. 

\item Contradictory to the reported deficit of a high-EW ($>$50\AA) LAE population at $z>7$ \citep[e.g.,][]{Tilvi2014a}, we find six LAEs with EW$>$50\AA, including one extremely large-EW (=280\AA) LAE. Along with other recent studies of finding high-EW LAEs \citep{Larson2018a, Jung2019a}, our result supports that a high-EW population is not extremely rare in the high-$z$ universe.

\item We estimate the asymmetry of Ly$\alpha$ emission lines with $\sigma_{\text{red}}/\sigma_{\text{blue}}$ from asymmetric Gaussian fitting. Our result reveals that an asymmetric profile of Ly$\alpha$ emission is common in the early universe, although the modest S/N of our emission lines results in low-significance asymmetry for most sources.

\item With the largest number of spectroscopic confirmations of galaxies at $z>7$ in a single study, we test the accuracy of their photometric redshifts, measuring the relative error of $\Delta z$. We notice a systematic bias at $z>7.4$ where photometric redshifts are always underestimated compared to the spectroscopic redshifts, although the overall quality of $z_{\text{phot}}$ appears good.  

\item We constrain the Ly$\alpha$ EW distribution at $7.0<z<8.2$, applying the methodology introduced in \cite{Jung2018a}, which constructs the PDF of the $e$-folding scale ($W_0$) of the EW distribution. Our constrained value of $W_0$ is $32^{+14}_{-9}$\AA\ at $z\sim7.6$, which is lower than lower-redshift values ($W_0\sim100$\AA). This implies an increasing \ion{H}{1} fraction at this redshift, although the derived $W_0$ values considerably depend on assumptions made about the intrinsic fraction of Ly$\alpha$ emitters among galaxies at different redshifts.

\item We study the $W_0$ dependence on UV magnitude with our statistical sample. Contradictory to the expectation from low-$z$ studies, which show a decreasing $W_0$ with an increasing UV-continuum brightness, we find that there is an apparent upturn of $W_0$ at the brightest objects at $z\sim7.6$, although the significance is low. This could be interpreted as a sign of different evolutions of Ly$\alpha$ EW between bright and faint objects during the epoch of reionization. 

\item We infer the IGM \ion{H}{1} fraction ($X_{\text{HI}}$) at $z\sim7.6$ based on our estimated Ly$\alpha$ transmission ($T^{\text{Ly$\alpha$}}_{\text{IGM}}$) in the IGM. Adopting a simplified analytical approach, our estimated $T^{\text{Ly$\alpha$}}_{\text{IGM}}$ corresponds to $X_{\text{HI}}\sim49^{+19}_{-19}\%$ at $z\sim7.6$. This is lower than the other recent measurements of Ly$\alpha$ spectroscopic surveys \citep{Hoag2019a, Mason2019a}, but close to the predicted value of \cite{Finkelstein2019b}, under the assumption that the ionizing photon budget from faint galaxies dominates.

\item A high Ly$\alpha$ detection rate at $z=7.5$ -- $7.6$, where we detect four Ly$\alpha$ emission lines, indicates an overdense and highly ionized region. Particularly, two pairs of Ly$\alpha$ emitters at $z=7.51$ and $z=7.60$ likely form localized ionized bubbles. These clustered LAEs could be a sign of a large (with a $\sim$ 40 cMpc spatial extent) highly ionized structure (or multiple smaller ionized bubbles) in this early universe. The existence of such ionized structures in our survey area could explain our lower inferred value of $X_{\text{HI}}$, though due to the expected inhomogeneity of reionization, such structures may be a common feature in this epoch.
\end{enumerate}

Recent measurements of the \ion{H}{1} fraction from Ly$\alpha$ surveys reported an extremely high \ion{H}{1} fraction ($X_{\text{HI}}$) at $z>7.5$: $X_{\text{HI}}$ = 0.88 at $z=7.6$, and $>$~0.76 at $z\sim8$ \citep{Mason2019a, Hoag2019a}. Although our estimation of $X_{\text{HI}}$ is based on a simplified analytic approach, our inferred $X_{\text{HI}}$ is below those high \ion{H}{1} fractions from other Ly$\alpha$ studies. However, it is consistent with the recent measurement of $X_{\text{HI}}=39^{+22}_{-13}\%$ from the damping wing feature of a luminous $z=7.5$ quasar observations in \cite{Yang2020a}, showing that a lower $X_{\text{HI}}$ fraction is plausible at $z >$ 7.

To resolve such tension between recent measurements, a wide-field Ly$\alpha$ spectroscopic survey is necessary to grasp the entire picture of reionization, overcoming cosmic variance, particularly toward the end of reionization as reionization is inhomogeneous. Our spectroscopic survey proves that we are able to detect Ly$\alpha$ with deep exposures even into the epoch of reionization, while most previous Ly$\alpha$ observations at this redshift are relatively shallow, resulting in lower detection rates. Thus, a direct measurement of the Ly$\alpha$ EW distribution over a wider area with enough sensitivity to detect Ly$\alpha$ will be necessary to allow us to capture a more comprehensive picture of reionization.

Our results put observational constraints on the redshift dependence of the Ly$\alpha$ EW distribution during the epoch of reionization, particularly toward the end of reionization, accounting for all forms of data incompleteness. However, constraining the \ion{H}{1} fraction in the IGM with Ly$\alpha$ is inevitably sensitive to reionization models \citep[e.g.,][]{Mesinger2015a, Kakiichi2016a, Mason2018a, Weinberger2019a}. Thus, implementing a realistic calculation of Ly$\alpha$ radiative processes \citep[e.g.,][]{Smith2019a, Kimm2019a} in future reionization models will place better predictions on how the expected Ly$\alpha$ EW distribution depends on the IGM neutral fraction. Moreover, the model predictions are dependent on many LAE systematics as well, such as the continuum luminosity, the interstellar medium (ISM) kinematics, and the stellar mass. For this purpose, a more comprehensive dataset covering various redshift ranges is required in order to investigate the Ly$\alpha$ systematics. 

With the arrival of powerful future telescopes, this will be facilitated with great ease. In upcoming years, the James Webb Space Telescope (JWST) will be capable of exploring this in depth. The JWST/NIRSpec will have a wide NIR wavelength coverage, probing other UV metal lines, including \ion{C}{3}], \ion{O}{2}], and \ion{O}{3}] lines for these high-$z$ galaxies. Utilizing other nonresonant metal lines allows us to derive the Ly$\alpha$ velocity offset with their systemic redshifts as shown in previous works \citep[e.g.,][]{Erb2014a, Steidel2016a, Stark2017a}. Investigating Ly$\alpha$ velocity offsets at high redshifts and comparing the velocity offsets between high- and low-$z$ LAE populations will provide a better understanding of the Ly$\alpha$ systematics. On the ground, the extremely large telescopes, such as the Giant Magellan Telescope (GMT) and the Thirty Meter Telescope (TMT), will play critical roles in exploring the reionization topology. Although JWST will probe key physical quantities of high-$z$ galaxies, a much wider field coverage of the GMT is necessary to grasp the entire picture of reionization, overcoming cosmic variance and capturing the inhomogeneous nature of reionization. Plus, the TMT with its planned NIR instrument, IRMS, will probe even fainter objects, utilizing its larger collecting area \cite[refer to][for more discussion on Ly$\alpha$ studies with the extremely large telescopes]{Finkelstein2019a}.

\acknowledgments
The authors would like to thank the anonymous referee for the detailed reading and the constructive suggestions on this paper. We thank Lennox Cowie, Charlotte Mason, and Kazuhiro Shimasaku for useful conversations.  I.J. acknowledges support by NASA under award number 80GSFC17M0002.  I.J. was supported from the NASA Headquarters under the NASA Earth and Space Science Fellowship Program--Grant 80NSSC17K0532. The authors wish to recognize and acknowledge the very significant cultural role and reverence that the summit of Maunakea has always had within the indigenous Hawaiian community. We are fortunate to have the opportunity to conduct our observations from this mountain. 

\appendix
\section{Summary of MOSFIRE Targets with Observed Spectral Energy Distributions}
We display the best-fit model SEDs of $z>6$ galaxies among entire sample in Figures \ref{fig:galaxy_seds1} and \ref{fig:galaxy_seds2}. All spectroscopic targets in our observations are list in Table \ref{tab:MOSFIREall}.

\startlongtable
\begin{deluxetable*}{ccccccccc}
\setlength{\tabcolsep}{12pt}
\tablecolumns{9}
\tablecaption{Summary of MOSFIRE Targets in GOODS-N\label{tab:MOSFIREall}}
\vspace{-5mm}
\tablehead{
\colhead{ID\tablenotemark{a}} & \colhead{R.A.}         & \colhead{Decl.}         & \colhead{$t_{\text{exp}}$} & \colhead{$J_{\text{125}}$} & \colhead{$M_{\text{UV}}$\tablenotemark{b}} & \colhead{$z_{\text{phot}}$\tablenotemark{c}} &\colhead{$z_{\text{spec}}$\tablenotemark{d}} & \colhead{EW$_{\text{Ly}\alpha}$\tablenotemark{e}}\\
\colhead{}     & \colhead{(J2000.0)} &\colhead{(J2000.0)} &  \colhead{(hrs)} & \colhead{} & \colhead{} &\colhead{} &\colhead{} & \colhead{(\AA)}
}
\startdata
{z7\_GND\_7157} & {189.225125} & { 62.286292} & { 5.8} & { 26.8} & {-20.5}& {7.54$^{+0.21}_{-5.82}$} &{8.128 (3.4)} &{21.2$^{+9.9}_{-8.0}$}\\
{z8\_GND\_8052} & {189.270000} & { 62.288558} & { 6.3} & { 26.5} & {-20.9}& {8.13$^{+0.27}_{-0.26}$} &{-} &{$<$42.1}\\
{z8\_GND\_41247} & {189.279500} & { 62.179753} & { 4.5} & { 27.1} & {-20.2}& {7.29$^{+0.46}_{-0.48}$} &{8.036 (3.9)} &{164.2$^{+85.8}_{-60.5}$}\\
{z7\_GND\_10402} & {189.179292} & { 62.275894} & {12.0} & { 25.5} & {-21.7}& {6.59$^{+0.06}_{-0.04}$} &{7.939 (4.0)} &{6.7$^{+2.7}_{-2.2}$}\\
{z7\_GND\_7831} & {189.177292} & { 62.291050} & { 4.5} & { 26.6} & {-20.6}& {7.89$^{+0.27}_{-0.24}$} &{-} &{$<$32.6}\\
{z7\_GND\_39781} & {189.251708} & { 62.185944} & { 4.5} & { 27.1} & {-20.1}& {6.94$^{+0.02}_{-0.03}$} &{7.881 (4.5)} &{123.9$^{+37.4}_{-32.9}$}\\
{z8\_GNW\_26779} & {189.286958} & { 62.318019} & { 5.5} & { 26.5} & {-20.6}& {7.83$^{+0.76}_{-0.32}$} &{-} &{$<$23.7}\\
{z7\_GND\_7376} & {189.243167} & { 62.285039} & { 5.8} & { 27.6} & {-19.5}& {6.45$^{+0.43}_{-0.77}$} &{7.768 (4.1)} &{32.5$^{+23.0}_{-13.0}$}\\
{z8\_GNW\_20826} & {189.401167} & { 62.319225} & { 7.2} & { 26.2} & {-21.1}& {7.73$^{+0.39}_{-0.37}$} &{-} &{$<$66.4}\\
{z7\_GND\_34204} & {189.359708} & { 62.205972} & { 4.5} & { 26.8} & {-20.4}& {7.08$^{+0.35}_{-0.29}$} &{7.608 (7.9)} &{279.7$^{+80.4}_{-62.5}$}\\
{z7\_GND\_16863} & {189.333083} & { 62.257236} & {16.2} & { 25.9} & {-21.2}& {7.21$^{+0.13}_{-0.13}$} &{7.599 (10.8)} &{61.3$^{+14.4}_{-11.4}$}\\
{z8\_GND\_21784} & {189.203125} & { 62.242486} & { 4.5} & { 26.5} & {-20.6}& {7.59$^{+0.24}_{-0.24}$} &{-} &{$<$30.8}\\
{z7\_GND\_6330} & {189.166250} & { 62.316497} & { 4.5} & { 26.4} & {-20.7}& {6.87$^{+0.14}_{-0.14}$} &{7.546 (6.1)} &{15.9$^{+4.4}_{-3.7}$}\\
{z8\_GND\_35384} & {189.232000} & { 62.202342} & { 4.5} & { 26.8} & {-20.2}& {7.51$^{+0.25}_{-0.27}$} &{-} &{$<$190.4}\\
{z7\_GND\_42912} & {189.157875} & { 62.302372} & {16.5} & { 25.5} & {-21.6}& {7.43$^{+0.11}_{-0.12}$} &{7.506 (10.8)} &{33.2$^{+4.3}_{-4.0}$}\\
{z7\_GNW\_32502} & {189.285833} & { 62.354964} & { 7.2} & { 26.6} & {-20.7}& {7.49$^{+0.64}_{-6.49}$} &{-} &{$<$93.1}\\
{z8\_GND\_7138} & {189.121208} & { 62.286400} & { 5.8} & { 27.3} & {-19.7}& {7.49$^{+0.33}_{-0.33}$} &{-} &{$<$33.1}\\
{z7\_GND\_18626} & {189.360667} & { 62.245567} & { 5.5} & { 27.5} & {-19.8}& {0.35$^{+6.61}_{-0.30}$} &{7.425 (4.6)} &{26.8$^{+14.9}_{-9.8}$}\\
{z8\_GND\_9408} & {189.300125} & { 62.280358} & {19.0} & { 27.2} & {-19.7}& {7.41$^{+0.44}_{-7.18}$} &{-} &{$<$39.4}\\
{z7\_GND\_42808} & {189.188417} & { 62.303050} & { 4.5} & { 26.5} & {-20.5}& {7.39$^{+0.15}_{-0.15}$} &{-} &{$<$36.1}\\
{z8\_GND\_22233} & {189.249792} & { 62.241225} & { 6.3} & { 26.3} & {-20.7}& {7.64$^{+0.11}_{-0.12}$} &{7.344 (7.1)} &{54.5$^{+15.0}_{-12.1}$}\\
{z7\_GND\_18323} & {189.371417} & { 62.252139} & {10.0} & { 26.2} & {-20.9}& {7.34$^{+0.21}_{-0.19}$} &{-} &{$<$19.4}\\
{z7\_GNW\_23317} & {189.439667} & { 62.302383} & { 7.2} & { 26.7} & {-20.5}& {7.31$^{+0.15}_{-0.15}$} &{-} &{$<$109.4}\\
{z8\_GND\_41470} & {189.224458} & { 62.311325} & {13.0} & { 25.9} & {-21.2}& {8.19$^{+0.08}_{-0.08}$} &{7.311 (3.5)} &{25.9$^{+9.5}_{-8.4}$}\\
{z7\_GNW\_19939} & {189.273375} & { 62.324783} & {12.7} & { 26.2} & {-20.9}& {7.27$^{+0.16}_{-0.17}$} &{-} &{$<$17.4}\\
{z7\_GND\_45190} & {189.138500} & { 62.275600} & {10.2} & { 26.7} & {-20.4}& {7.38$^{+0.24}_{-0.23}$} &{7.265 (3.4)} &{22.9$^{+13.7}_{-9.3}$}\\
{z7\_GNW\_32653} & {189.278750} & { 62.357453} & { 7.2} & { 26.2} & {-21.0}& {7.26$^{+0.10}_{-0.12}$} &{-} &{$<$61.7}\\
{z7\_GND\_6451} & {189.222000} & { 62.315761} & {12.0} & { 27.0} & {-20.0}& {7.24$^{+0.14}_{-0.15}$} &{7.246 (3.2)} &{43.2$^{+17.2}_{-15.1}$}\\
{z7\_GNW\_18773} & {189.309167} & { 62.362142} & { 7.2} & { 26.4} & {-20.6}& {7.22$^{+0.30}_{-0.22}$} &{-} &{$<$78.5}\\
{z7\_GND\_11402} & {189.186167} & { 62.270864} & {12.0} & { 25.4} & {-21.6}& {7.22$^{+0.08}_{-0.08}$} &{-} &{$<$5.8}\\
{z7\_GND\_13456} & {189.264958} & { 62.265789} & {11.8} & { 26.1} & {-20.9}& {7.16$^{+0.12}_{-0.12}$} &{-} &{$<$12.7}\\
{z7\_GND\_6739} & {189.260417} & { 62.289403} & { 5.8} & { 26.9} & {-20.1}& {7.14$^{+0.23}_{-0.29}$} &{-} &{$<$26.6}\\
{z7\_GND\_44088} & {189.135000} & { 62.291869} & {10.7} & { 27.1} & {-19.9}& {7.59$^{+0.21}_{-0.20}$} &{7.133 (5.2)} &{87.6$^{+23.8}_{-21.2}$}\\
{z7\_GNW\_21903} & {189.290750} & { 62.311100} & {11.8} & { 27.1} & {-20.3}& {7.10$^{+5.26}_{-4.99}$} &{-} &{$<$27.8}\\
{z7\_GND\_13934} & {189.275917} & { 62.260303} & { 5.5} & { 27.1} & {-19.9}& {7.09$^{+0.16}_{-0.17}$} &{-} &{$<$41.4}\\
{z7\_GNW\_30165} & {189.327958} & { 62.298572} & { 5.5} & { 26.3} & {-20.6}& {7.07$^{+0.40}_{-6.44}$} &{-} &{$<$15.7}\\
{z8\_GND\_21308} & {189.109500} & { 62.238792} & { 5.8} & { 26.8} & {-20.2}& {7.06$^{+0.17}_{-0.12}$} &{-} &{$<$17.9}\\
{z7\_GNW\_18834} & {189.273292} & { 62.360786} & { 7.2} & { 25.8} & {-21.1}& {7.05$^{+0.21}_{-0.21}$} &{-} &{$<$46.3}\\
{z7\_GND\_15330} & {189.310875} & { 62.260447} & {11.8} & { 27.2} & {-19.7}& {7.03$^{+0.14}_{-0.13}$} &{-} &{$<$36.4}\\
{z7\_GND\_21360} & {189.299458} & { 62.236889} & { 5.5} & { 27.0} & {-20.0}& {6.97$^{+0.11}_{-0.10}$} &{-} &{$<$31.7}\\
{z7\_GND\_18869} & {189.205292} & { 62.250767} & {16.5} & { 26.4} & {-20.4}& {6.96$^{+0.02}_{-0.05}$} &{-} &{$<$11.9}\\
{z7\_GND\_16441} & {189.082667} & { 62.252475} & { 5.8} & { 25.8} & {-21.2}& {6.96$^{+0.12}_{-0.11}$} &{-} &{$<$8.2}\\
{z8\_GNW\_23312} & {189.166250} & { 62.316497} & { 6.3} & { 26.4} & {-20.6}& {6.87$^{+0.14}_{-0.14}$} &{-} &{$<$24.6}\\
{z7\_GNW\_22375} & {189.166250} & { 62.316497} & {19.0} & { 26.4} & {-20.6}& {6.87$^{+0.14}_{-0.14}$} &{-} &{$<$17.9}\\
{z7\_GND\_40057} & {189.339458} & { 62.184781} & { 4.5} & { 26.7} & {-20.2}& {6.85$^{+0.17}_{-0.16}$} &{-} &{$<$136.3}\\
{z8\_GND\_24214} & {189.201042} & { 62.227439} & { 5.8} & { 27.1} & {-19.8}& {6.85$^{+0.07}_{-0.09}$} &{-} &{$<$26.2}\\
{z7\_GND\_11368} & {189.283708} & { 62.272244} & {11.8} & { 26.9} & {-19.8}& {6.82$^{+0.18}_{-0.18}$} &{-} &{$<$48.1}\\
{z7\_GNW\_24671} & {189.361708} & { 62.294372} & { 7.2} & { 25.9} & {-21.1}& {6.81$^{+0.05}_{-0.07}$} &{-} &{$<$45.5}\\
{z7\_GNW\_28411} & {189.392708} & { 62.308617} & { 5.5} & { 25.9} & {-20.8}& {6.78$^{+0.11}_{-0.18}$} &{-} &{$<$15.0}\\
{z7\_GND\_38350} & {189.177167} & { 62.291519} & { 5.8} & { 26.2} & {-20.8}& {6.77$^{+0.06}_{-0.06}$} &{-} &{$<$16.3}\\
{z7\_GND\_35507} & {189.304458} & { 62.201678} & { 4.5} & { 26.7} & {-20.2}& {6.74$^{+0.07}_{-0.07}$} &{-} &{$<$122.0}\\
{z7\_GNW\_30851} & {189.356875} & { 62.295319} & { 5.5} & { 24.7} & {-22.0}& {6.73$^{+0.04}_{-0.04}$} &{-} &{$<$4.6}\\
{z7\_GND\_15163} & {189.079833} & { 62.256458} & { 5.8} & { 25.6} & {-21.3}& {6.72$^{+0.12}_{-0.11}$} &{-} &{$<$7.2}\\
{z7\_GND\_4369} & {189.187333} & { 62.318822} & { 5.8} & { 27.3} & {-19.7}& {6.70$^{+0.18}_{-0.17}$} &{-} &{$<$24.6}\\
{z7\_GND\_38613} & {189.155292} & { 62.286461} & { 5.8} & { 26.6} & {-20.3}& {6.68$^{+0.05}_{-0.05}$} &{-} &{$<$17.8}\\
{z7\_GND\_36688\tablenotemark{f}} & {189.178125} & { 62.310639} & { 5.8} & { 27.4} & {-}& {6.64$^{+0.23}_{-0.23}$} &{-} &{-}\\
{z7\_GND\_38200} & {189.359167} & { 62.192122} & { 4.5} & { 26.6} & {-20.1}& {6.56$^{+0.28}_{-5.08}$} &{-} &{$<$157.6}\\
{z7\_GND\_43951} & {189.154625} & { 62.292919} & {12.0} & { 27.3} & {-19.5}& {6.55$^{+0.11}_{-0.11}$} &{-} &{$<$30.1}\\
{z7\_GND\_15642} & {189.151958} & { 62.259639} & {12.0} & { 26.8} & {-20.1}& {6.51$^{+0.21}_{-0.20}$} &{-} &{$<$21.2}\\
{z7\_GND\_22525} & {189.372833} & { 62.240372} & {10.0} & { 27.6} & {-19.2}& {6.48$^{+0.47}_{-0.53}$} &{-} &{$<$73.1}\\
{z7\_GND\_43678\tablenotemark{f}} & {189.235208} & { 62.295594} & { 6.3} & { 26.5} & {-}& {6.40$^{+0.08}_{-0.09}$} &{-} &{-}\\
{z6\_GND\_11714} & {189.323917} & { 62.271114} & { 6.3} & { 28.2} & {-18.7}& {6.10$^{+0.25}_{-0.27}$} &{-} &{$<$105.1}\\
{z6\_GNW\_20715} & {189.339875} & { 62.319981} & { 7.2} & { 25.9} & {-}& {5.54$^{+0.07}_{-0.07}$} &{-} &{-}\\
{z7\_GND\_22483} & {189.146417} & { 62.240519} & {16.5} & { 26.9} & {-}& {4.51$^{+0.16}_{-0.26}$} &{-} &{-}\\
{z6\_GND\_12175} & {189.172042} & { 62.269725} & { 6.3} & { 28.2} & {-}& {1.88$^{+1.15}_{-1.39}$} &{-} &{-}\\
{z7\_GND\_22782} & {189.353000} & { 62.232217} & { 5.5} & { 25.2} & {-}& {1.55$^{+0.08}_{-0.08}$} &{-} &{-}\\
{z7\_GND\_25452} & {189.136583} & { 62.223317} & { 5.8} & { 25.2} & {-}& {1.53$^{+0.05}_{-0.05}$} &{-} &{-}\\
{z8\_GND\_7253} & {189.264250} & { 62.285675} & { 5.5} & { 27.1} & {-}& {1.30$^{+0.29}_{-0.10}$} &{-} &{-}\\
{z6\_GND\_41772} & {189.161083} & { 62.309928} & { 6.3} & { 27.2} & {-}& {1.14$^{+4.38}_{-0.15}$} &{-} &{-}\\
{z6\_GND\_39946} & {189.319583} & { 62.185250} & { 4.5} & { 26.7} & {-}& {1.02$^{+4.53}_{-0.24}$} &{-} &{-}\\
{z8\_GNW\_20236} & {189.433500} & { 62.322303} & { 7.2} & { 26.1} & {-}& {0.69$^{+6.09}_{-0.02}$} &{-} &{-}\\
{z7\_GND\_11273} & {189.100375} & { 62.268478} & { 5.8} & { 27.3} & {-}& {0.38$^{+0.38}_{-0.32}$} &{-} &{-}\\
\enddata
\tablecomments{\\$^{a}$\footnotesize The listed IDs are from \cite{Finkelstein2015a}, encoded with their photometric redshifts and the fields in the CANDELS imaging data.\\
$^{b}$$M_{\text{UV}}$ is estimated from the averaged flux over a 1450 -- 1550\AA\ bandpass from the best-fit galaxy SED model.\\
$^{c}$We present the 1$\sigma$ range of  $z_{\text{phot}}$.\\
$^{d}$Spectroscopic redshifts are estimated from the detected Ly$\alpha$ emission line, and the values in parentheses are their S/N ratios.\\
$^{e}$$3\sigma$ upper limits for non-detection objects, measured from the median flux limits between sky-emission lines from individual spectra. Also we do not display EW upper limits for galaxies with 2$\sigma$ upper limits of $z_{\text{phot}}<6$, which are listed at the bottom of the table.\\
$^{f}$Emission line detected, but likely to be a low-$z$ object.}
\end{deluxetable*}

\begin{figure*}[ht]
\centering
\includegraphics[width=0.84\paperwidth]{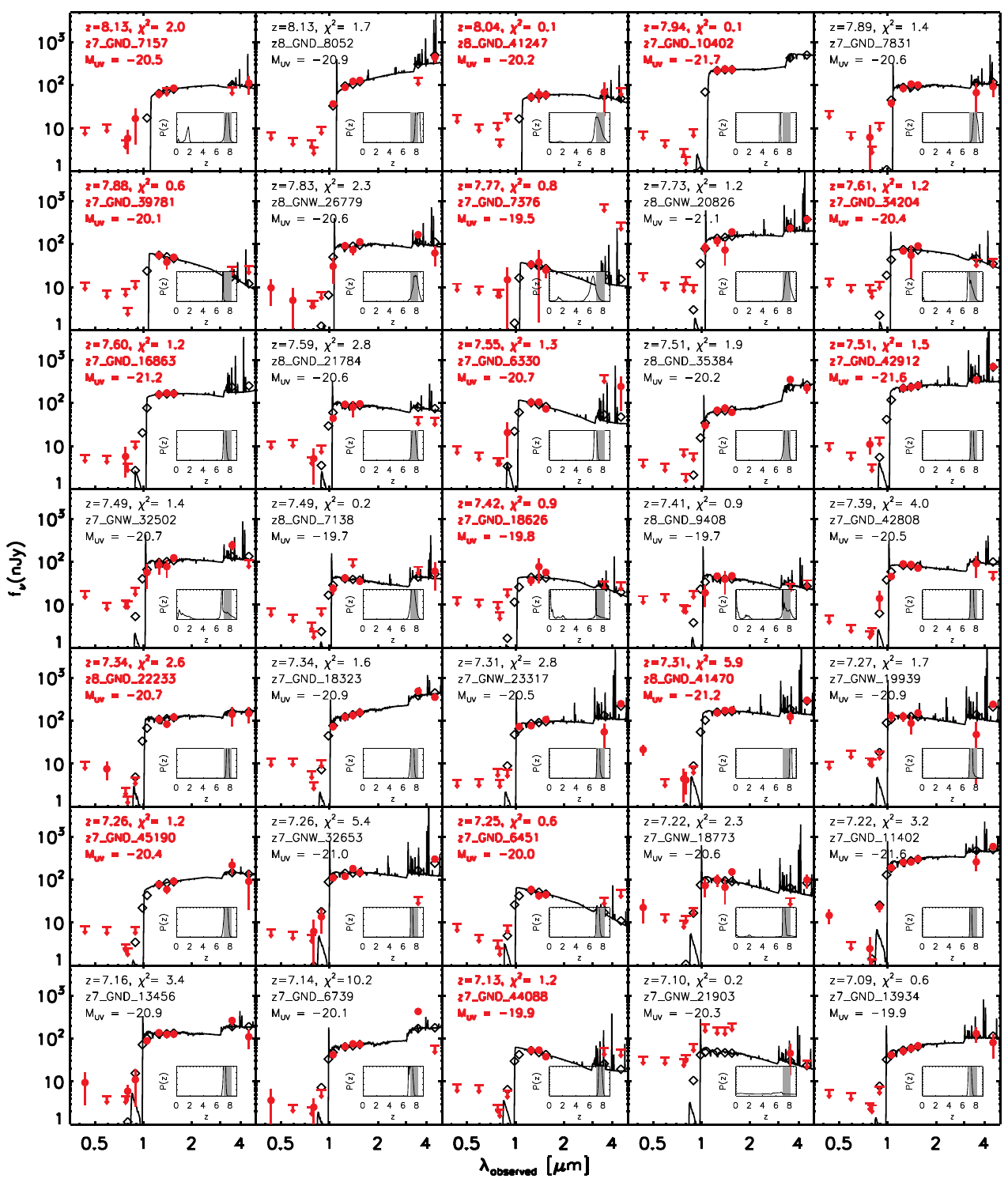}
\caption{The best-fit models SEDs of $z>6$ galaxies among the entire sample. Each panel displays the best-fit model (solid curves and diamond symbols) with the observed photometry (red filled circles). Downward arrows denote 1$\sigma$ upper limits. The panels with red bold texts display Ly$\alpha$-detected galaxies. For Ly$\alpha$ non-detected galaxies, we assume galaxy redshifts as the peak redshift values of the photo-$z$ PDFs.  The normalized photo-$z$ PDFs are shown in inset panels, and the shaded regions denote the instrumental redshift coverage of Ly$\alpha$ emission in MOSFIRE $Y$-band observations.} 
\label{fig:galaxy_seds1}
\end{figure*}

\begin{figure*}[ht]
\centering
\includegraphics[width=0.84\paperwidth]{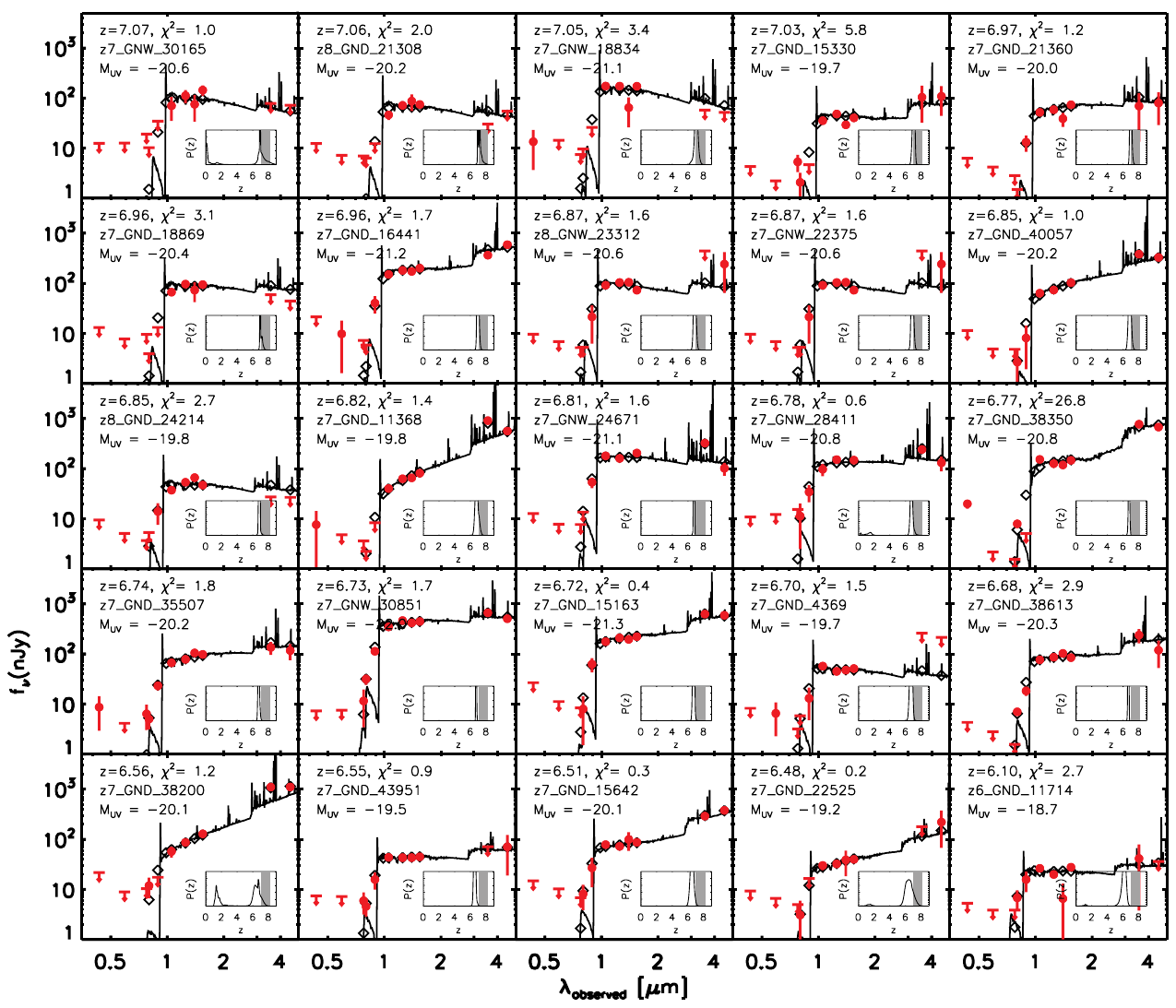}
\caption{Continued from previous page.} 
\label{fig:galaxy_seds2}
\end{figure*}

\bibliographystyle{aasjournal}
\scriptsize
\bibliography{references}
\end{document}